\documentclass{aastex63}

\newcommand{\mh}{\mbox{\rm [{\rm M}/{\rm H}]}}
\newcommand{\Msun}{\mbox{$M_{\odot}$}}

\newcommand{\beq}{\begin{equation}}
\newcommand{\eeq}{\end{equation}}
\newcommand{\beqa}{\begin{eqnarray}}
\newcommand{\eeqa}{\end{eqnarray}}
\newcommand{\Rap}{\mbox{$R_{\rm ap}$}}
\newcommand{\Rsc}{\mbox{$R_{\rm sc}$}}
\newcommand{\fB}{\mbox{${\rm F475W}$}}
\newcommand{\fI}{\mbox{${\rm F814W}$}}
\newcommand{\fJ}{\mbox{${\rm F110W}$}}
\newcommand{\fH}{\mbox{${\rm F160W}$}}
\newcommand{\fBI}{\mbox{$\fB\!-\!\fI$}}
\newcommand{\fJH}{\mbox{$\fJ\!-\!\fH$}}

\newcommand{\fcrowd}{\mbox{${\rm F160W}_{\rm crowd}$}}

\newcommand{\comment}[1]{}

\received{\today}
\revised{XXX}
\accepted{XXX}
\submitjournal{ApJ}

\shorttitle{AGBs and cool giants in M31 clusters}
\shortauthors{Girardi et al.}

\begin{document}

\title{PHAT XX. AGB stars and other cool giants in M31 star clusters}

\correspondingauthor{L\'eo Girardi}

\email{leo.girardi@inaf.it}

\author[0000-0002-6301-3269]{L\'eo Girardi}
\affiliation{Osservatorio Astronomico di Padova -- INAF, Vicolo dell'Osservatorio 5, I-35122 Padova, Italy}
\author[0000-0003-4850-9589]{Martha L. Boyer}
\affiliation{Space Telescope Science Institute, 3700 San Martin Drive, Baltimore, MD 21218, USA}
\author[0000-0001-6421-0953]{L.~Clifton Johnson} 
\affiliation{Department of Physics and Astronomy, Northwestern University, 2145 Sheridan Road, Evanston, IL 60208, USA}
\author[0000-0002-1264-2006]{Julianne J. Dalcanton}
\affiliation{Department of Astronomy, University of Washington, Box 351580, Seattle, WA 98195, USA}
\author[0000-0001-9306-6049]{Philip Rosenfield} 
\affiliation{Eureka Scientific, Inc., 2452 Delmer Street, Oakland, CA 94602, USA}
\author[0000-0003-0248-5470]{Anil C. Seth}
\affiliation{Department of Physics and Astronomy, University of Utah, Salt Lake City, UT 84112, USA}
\author[0000-0003-0605-8732]{Evan D. Skillman}
\affiliation{Minnesota Institute for Astrophysics, University of Minnesota, 116 Church St. SE, Minneapolis, MN 55455}
\author[0000-0002-6442-6030]{Daniel R. Weisz}
\affiliation{Department of Astronomy, University of California, 501 Campbell Hall \#3411, Berkeley, CA 94720-3411, USA}
\author[0000-0002-7502-0597]{Benjamin F. Williams}
\affiliation{Department of Astronomy, University of Washington, Box 351580, Seattle, WA 98195, USA}
\author[0000-0002-4342-4626]{Antara Raaghavi Bhattacharya}
\affiliation{Navy Children School, Mumbai, India}
\affiliation{Science Internship Program (SIP) intern, University of California Santa Cruz, Department of Astronomy and Astrophysics, 1156 High Street, Santa Cruz, California 95064, USA}
\author[0000-0002-7922-8440]{Alessandro Bressan}
\affiliation{SISSA, via Bonomea 365, I-34136 Trieste, Italy} 
\author[0000-0003-2352-3202]{Nelson Caldwell}
\affiliation{Harvard-Smithsonian Center for Astrophysics, 60 Garden Street, Cambridge, MA 02138, USA}, 
\author[0000-0002-3759-1487]{Yang Chen}
\affiliation{Dipartimento di Fisica e Astronomia Galileo Galilei, Universit\`a di Padova, Vicolo dell'Osservatorio 3, I-35122 Padova, Italy}
\author[0000-0001-8416-4093]{Andrew E. Dolphin}
\affiliation{Raytheon, 1151 E. Hermans Road, Tucson, AZ 85706, USA}
\affiliation{Steward Observatory, University of Arizona, 933 North Cherry Avenue, Tucson, AZ 85721, USA} 
\author[0000-0001-9256-5516]{Morgan Fouesneau}
\affiliation{Max-Planck-Institut f\"ur Astronomie, K\"onigstuhl 17, D-69117 Heidelberg, Germany}
\author[0000-0002-8937-3844]{Steven Goldman}
\affiliation{Space Telescope Science Institute, 3700 San Martin Drive, Baltimore, MD 21218, USA}
\author[0000-0001-8867-4234]{Puragra Guhathakurta}
\affiliation{UCO/Lick Observatory, University of California Santa Cruz, 1156 High Street, Santa Cruz, CA 95064, USA}
\author[0000-0002-9137-0773]{Paola Marigo}
\affiliation{Dipartimento di Fisica e Astronomia Galileo Galilei, Universit\`a di Padova, Vicolo dell'Osservatorio 3, I-35122 Padova, Italy}
\author[0000-0003-1622-1302]{Sagnick Mukherjee}
\affiliation{UCO/Lick Observatory, University of California Santa Cruz, 1156 High Street, Santa Cruz, CA 95064, USA}
\affiliation{Department of Physics, Presidency University, Kolkata, India}
\author[0000-0002-9300-7409]{Giada Pastorelli}
\affiliation{Space Telescope Science Institute, 3700 San Martin Drive, Baltimore, MD 21218, USA}
\author[0000-0001-8481-2660]{Amanda Quirk}
\affiliation{UCO/Lick Observatory, University of California Santa Cruz, 1156 High Street, Santa Cruz, CA 95064, USA}
\author[0000-0001-6360-992X]{Monika Soraisam}
\affiliation{National Center for Supercomputing Applications, University of Illinois at Urbana-Champaign, Urbana, IL 61801, USA}
\affiliation{Department of Astronomy, University of Illinois at Urbana-Champaign, Urbana, IL 61801, USA}
\author[0000-0002-1429-2388]{Michele Trabucchi}
\affiliation{Department of Astronomy, University of Geneva, Ch. des Maillettes 51, 1290 Versoix, Switzerland}
\affiliation{Dipartimento di Fisica e Astronomia Galileo Galilei, Universit\`a di Padova, Vicolo dell'Osservatorio 3, I-35122 Padova, Italy}




\begin{abstract}
The presence of AGB stars in clusters provides key constraints for stellar models, as has been demonstrated with historical data from the Magellanic Clouds. In this work, we look for candidate AGB stars in M31 star clusters from the Panchromatic Hubble Andromeda Treasury (PHAT) survey. Our photometric criteria selects stars brighter than the tip of the red giant branch, which includes the bulk of the thermally-pulsing AGB stars as well as early-AGB stars and other luminous cool giants expected in young stellar populations (e.g. massive red supergiants, and intermediate-mass red helium-burning stars). The AGB stars can be differentiated, a posteriori, using the ages already estimated for our cluster sample. 937 candidates are found within the cluster aperture radii, half ($\sim450$) of which are very likely cluster members. Cross-matching with additional databases reveals two carbon stars and ten secure variables among them. The field-corrected age distribution reveals the presence of young supergiants peaking at ages smaller than $10^8$~yr, followed by a long tail of AGB stars extending up to the oldest possible ages. This long tail reveals the general decrease in the numbers of AGB stars from initial values of $\sim50\times10^{-6}\Msun^{-1}$ at $10^8$~yr down to $\sim5\times10^{-6}\Msun^{-1}$ at $10^{10}$~yr. Theoretical models of near-solar metallicity reproduce this general trend, although with localized discrepancies over some age intervals, whose origin is not yet identified. The entire catalogue is released together with finding charts to facilitate follow-up studies. 
\end{abstract}




\section{Introduction}
\label{intro}

Stellar clusters with known ages and metallicities allow one to securely anchor the initial masses of stars evolving within them. This approach is particularly valuable for stars that are intrinsically challenging to model, due to their rapid evolution and complex interior physics. Asymptotic giant branch (AGB) stars, in particular, have benefited from systematic studies of their properties in stellar clusters, although such studies have been limited due to their rarity in any individual cluster.

There are at least 124 AGB stars belonging to 31 Magellanic Clouds' (MC) clusters \citep{Frogel_etal90}. Some remarkable cases are NGC~419 in the SMC and NGC~1846 in the LMC, with $\sim20$ likely AGB members each, which are probably the places in the known universe with the highest spatial concentration of resolved carbon stars. 

During the dawn of near-infrared astronomy, in the 1970s and 1980s, observations of these Magellanic Cloud clusters have been crucial to the development of a general picture about the evolution of AGB stars, especially regarding their final thermally-pulsing (TP-AGB) phase where the processes of third dredge-up, hot bottom-burning, long-period variability and dust-driven mass loss take place \citep[see][for reviews]{lattanzio04,herwig05,hoefner18}. By the time of \citet{Frogel_etal90}, some basic facts were already clear, for instance
\begin{enumerate}
    \item that the initially more massive stars, found in the younger star clusters, reach higher luminosities along the TP-AGB than less massive stars, found in the oldest clusters;
    \item that luminous giants rich in C-bearing molecules (hereafter C stars) are generated only inside a limited range of initial masses, in contrast with the O-rich giants (generally known as M giants);
    \item that the TP-AGB lifetimes, limited to luminosities above the tip of the RGB of intermediate-age and old populations, are of the order of a few Myr at most.
\end{enumerate}

Subsequent works used these cluster data to derive more quantitative constraints on several aspects of the AGB evolution, including those on the C and M type lifetimes \citep{GirardiMarigo07}, on the fraction of the integrated light contributed by the TP-AGB phase \citep{Maraston05, Pessev_etal08, Noel_etal13}, on the dust emission by the mass-losing, extreme AGB stars \citep{vanLoon_etal05}, and on the connections between mass loss, long-period variability and surface chemical composition \citep{Lebzelter_etal08, Kamath_etal10, Kamath_etal12}. 

Yet despite their richness, relative proximity, little reddening, and nearly constant distance, the MC clusters are far from being the ideal sample for the study of AGB stars. They do not uniformly cover the age interval in which AGB stars are produced \citep{Marigo_etal96}, and span a limited range of metallicities. Moreover, even for the richest MC clusters, there are significant stochastic fluctuations in the numbers of evolved stars, which is reflected in all quantities from them derived, including their integrated colors \citep{SantosFrogel97}. 

In principle, some of these problems can be alleviated by grouping several clusters into a few age bins \citep[as done by e.g.][]{GirardiMarigo07, Maraston05, Pessev_etal08, Noel_etal13}. However, some of the quantitative constraints on the AGB evolution derived in this way have been questioned after the discovery of the ``AGB boosting'' effect at ages of $\sim\!1.6$~Gyr \citep{Girardi_etal13}: In short, it happens that for a limited interval of ages the numbers of observed AGB stars in clusters (and all derived quantities) are not proportional to the AGB lifetime -- and, unfortunately, most of the MC cluster data concentrates close to this age limit.

That said, it would be very important if other rich samples of AGB stars belonging to star clusters were available, so that the trends observed in the MCs could be checked and extended to a wider range of ages and metallicities. Star clusters in the Milky Way galaxy (MW) are not that useful in this regard. Although many MW old globular clusters have a clear ``early-AGB bump'' along their RGBs \citep[see e.g.][]{ferraro99}, they generally do not present many stars above the TRGB that could be clearly assigned to the TP-AGB phase, with a few exceptions like 47~Tuc \citep{LebzelterWood05, Lebzelter_etal14,mcdonald11,momany12} and {$\omega$}~Cen \citep{Boyer_etal08, McDonald_etal09, McDonald_etal11}. In many other cases the MW globular clusters host just a handful of candidate TP-AGB stars selected on the basis of their long-period variability \citep{LebzelterWood11}. Nearby open clusters are in general too poorly populated to present even a single AGB star. Star clusters in other Local Group dwarf galaxies suffer from the same problems. Beyond the Local Group, there are many super-star clusters that probably contain many AGB stars each  \citep[e.g.][]{schweizer96, whitmore99}. They however are not resolved into individual objects even when observed with the Hubble Space Telescope (HST). 

The Andromeda (M31) and Triangulum (M33) galaxies are the obvious candidates to obviate such a situation. They are located at distances in which they are resolvable into stars by the HST, contain stellar populations of all ages \citep{ferguson05, brown06, Javadi_etal11, Bernard_etal12, Bernard_etal14, lewis15, williams17}, and host hundreds of candidate star clusters \citep{bolognacat, sarajedini07, sanroman10, peacock10, Johnson_etal12}. Until recently, however, the necessary HST imaging was not available, beyond limited areas across these galaxies and for a few of its most conspicuous clusters \citep[e.g.][]{williams01, caldwell09}. More critically, the available high-resolution imaging did not include the near-infrared (NIR), which is crucial for a clear identification of the TP-AGB stars. 

For M31, the situation has radically changed with the Panchromatic Hubble Andromeda Treasury \citep[PHAT;][]{Dalcanton_etal12} survey, which collected UV, optical, and NIR imaging for about 1/3 of M31 disk, and most of its bulge. PHAT now includes $2753$ star clusters identified through the citizen-scientists Andromeda Project\footnote{\url{http://www.andromedaproject.org/}} \citep[AP;][]{johnson15}. They are a rich ground for searches of stars in relatively rapid phases of their evolution, as demonstrated by the identification of a number of Cepheids \citep{senchyna15} and a planetary nebula \citep{davis19} in them. It is also interesting to note that the PHAT mean spatial resolution (1 ACS/WFC pixel = 0.19 pc in the optical, 1 WFC3/IR pixel = 0.5 pc in the NIR) is similar to the resolution which was obtained from the ground for MC clusters (1\arcsec = 0.25 pc for the LMC) at the dawn of NIR astronomy \citep[][and references therein]{Frogel_etal90}, and is still comparable to the resolution being obtained in present studies of the MCs in the NIR \citep[e.g.,][]{Cioni_etal11}. 


In this paper, we provide a first list of AGB candidates in M31 star clusters, and discuss their reliability. We look for candidates at luminosities above the TRGB, which means that the sample may include multiple stellar types, including non-AGB stars at young ages (see discussion in Sect.~\ref{results}), a combination of early-AGB and TP-AGB stars at intermediate-ages, then exclusively TP-AGB stars at ages older than $\sim3$~Gyr. Throughout this work, we will use a generic ``AGB'' label to refer to the cases in which AGB stars (both early-AGB and TP-AGB) are likely present, and an even more generic ``AGB candidate'' to refer to the samples which may include a significant fraction of very young, non-AGB stars. 
Data and methods are described in Sect.~\ref{data}. The properties of the derived sample are discussed in Sect.~\ref{results}. A few conclusions are drawn in Sect.~\ref{conclu}.

\section{Data and methods}
\label{data}

\subsection{PHAT imaging and photometry}

\begin{figure}
\includegraphics[width=0.49\textwidth]{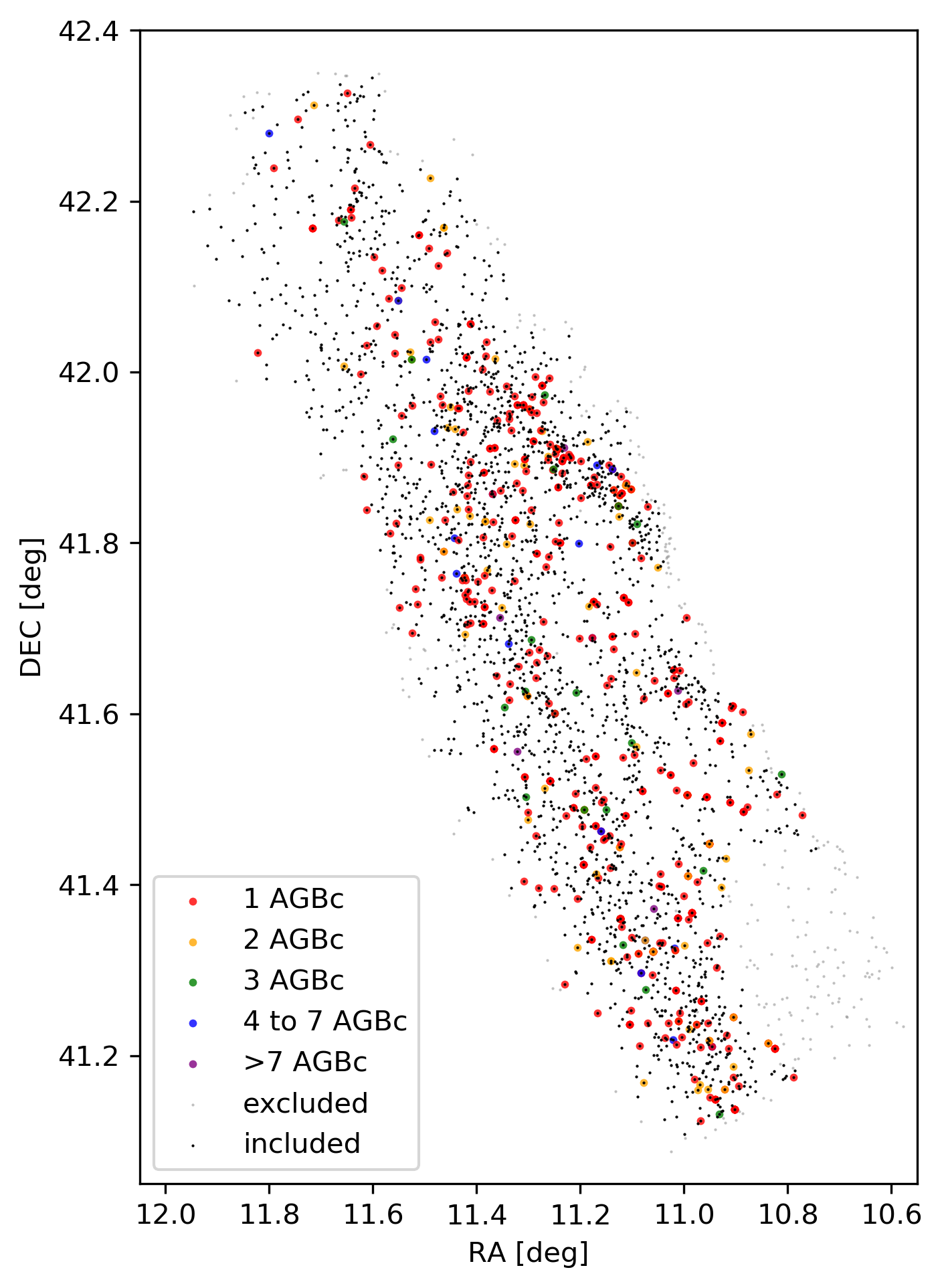}
\caption{
Sky distribution of PHAT star clusters from the AP catalog (small dots). The gray dots mark clusters excluded from our sample, either because they belong to the M31 bulge (at the bottom-right corner) or because they do not have NIR photometry. The remaining black dots are clusters searched for AGB stars. The coloured circles mark the clusters where they are found, with their numbers per cluster as in the legend. 
}
\label{fig_footprint}
\end{figure}

The PHAT survey is extensively described in \citet{Dalcanton_etal12}. Its extended, 0.5-deg$^2$ wide footprint can be appreciated in Fig.~\ref{fig_footprint}. The HST imaging was organized into 23 ``bricks'' of $6\arcmin\times12\arcmin$ each, formed by $3\times6$ contiguous pointings (or ``fields'') of the Wide Field Camera 3 (WFC3) near-infrared channel. The same bricks were imaged in two ultraviolet bands with the WFC3 ultraviolet/optical (UVIS) channel, and at a separation of 6 months in the optical with the Advanced Camera for Surveys (ACS) Wide Field Channel (WFC). \citet{williams14} describe the complex work of aligning all the images and performing simultaneous 6-filter photometry in them. It resulted in photometric measurements for over 117 million stars in M31. The final catalogue is available at the Mikulski Archive for Space Telescopes (MAST)\footnote{\url{https://archive.stsci.edu/prepds/phat/}}, and contains several useful photometry quality flags. 

The best photometry is obtained for the two optical filters of ACS/WFC, F475W and F814W, which present the most favourable combination of exposure time and spatial resolution. Indeed, in these filters the photometry reaches more than 3~mag below the red clump, except in the most crowded areas close to the M31 bulge. The WFC3/IR photometry in F110W and F160W is also of excellent quality, thanks to the lower level of crowding caused by the blue main sequence stars at these wavelengths and by the use of simultaneous optical and IR photometry \citep{williams14}. The NIR photometry is generally limited to the upper RGB and above, but also extends to below the red clump in the outermost PHAT bricks. 

\subsection{Cluster catalogues}

In this paper we use the AP cluster catalogue from \citet{johnson15}. It comprises the complete PHAT area as illustrated in Fig.~\ref{fig_footprint}. It includes 2753 clusters identified by citizen scientists and later verified with the aid of the Year 1 catalogue \citep[a dataset identified by professional astronomers, limited to 4 PHAT bricks plus 2 half-bricks, see][]{Johnson_etal12} and with fake clusters inserted in the original images. 
Because the catalog is sorted with the highest fractions of users identifying clusters first, objects with smaller AP numbers correspond to clusters more evident to the eye -- which usually means they are younger, more populous, and in M31 areas with a simpler foreground/background. Clusters in the AP catalog are also characterized by their central coordinates, and by their ``aperture radii'', \Rap, which mark the visible extent of the clusters. They are calculated using the median values of the individual cluster centres and radii measured by AP users. 

From the AP catalogue, we have excluded clusters without NIR imaging from PHAT, and 106 clusters belonging to the highly-crowded bulge area (see figure 12 in \citealt{johnson15}). We are then left with a sample of 2501 clusters. Physical parameters of the clusters have been determined in an homogeneous way by \citet{johnson15}. The cluster ages include color-magnitude ages for younger clusters ($\la300$~Myr) as presented in \citet{johnson16}, and integrated light estimates of the ages and masses using \citet{Fouesneau_etal14}'s method by \citet{beerman15}.

Given the way the AP catalogue was built and the variety of cluster sizes, concentrations, and foreground/background across the M31 disk, there is no one-to-one relationship between \Rap\ and other commonly-adopted cluster structural parameters such as the half-light radius, $R_\mathrm{eff}$. But cluster-to-cluster estimates of $R_\mathrm{eff}$ based on radial profiles \citep[table 2 in][]{johnson15}, and extensive experiments with fake clusters added to PHAT images (their table 6), indicate that \Rap\ is typically between 1 and 5 times larger than $R_\mathrm{eff}$. Just 3~\% of the fake clusters presented by \citet{johnson15} turn out to have \Rap\ smaller than $R_\mathrm{eff}$.

\subsection{Candidate AGB stars in PHAT}

\begin{figure}
\includegraphics[width=0.49\textwidth]{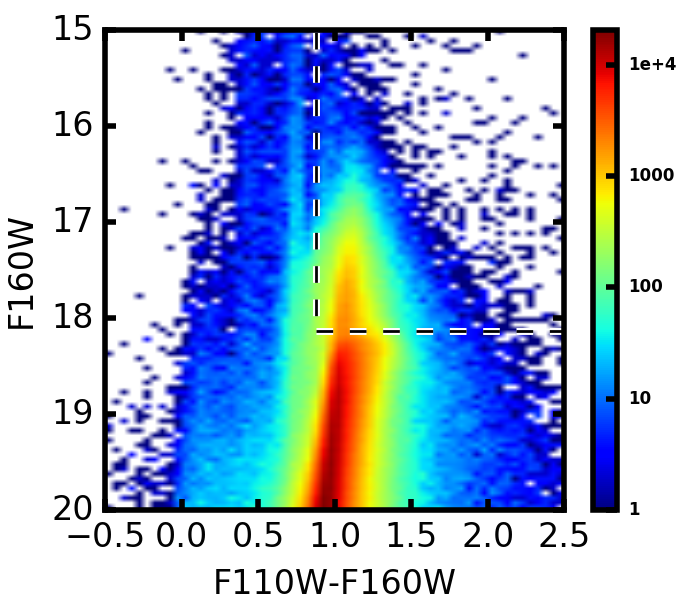}
\caption{Field NIR CMD from the PHAT low-density regions (density of RGB stars with $18.5<\mathrm{F160W}<19.5$ smaller than 0.3 arcsec$^{-2}$, see \citealt{williams14}), illustrating the selection of AGB candidates in the region above and to the right of the dashed line. The selected stars are just red enough to avoid the vertical stripe of foreground low-mass main sequence stars at $\mathrm{F110W}-\mathrm{F160W} \sim\!0.7$~mag, and just bright enough to avoid the well-populated RGB region starting at $\fH\ga18.2$~mag.}
\label{fig_cmd}
\end{figure}

From the 6-filter photometry of low-surface-brightness areas, we build the NIR CMD of Fig.~\ref{fig_cmd}. The sample is limited to stars that satisfy the ``GST'' quality criterion for the photometry defined in \citet{williams14}, in both NIR passbands. In short, the GST criterion is designed to eliminate objects strongly affected by blending, cosmic rays, or instrument artifacts. In the NIR filters, it consists in keeping only measurements for which the photometric pipeline \citep[namely DOLPHOT; see][]{dolphot} returned a signal-to-noise larger than 4, a sharpness-squared smaller than 0.15, and a crowding parameter smaller than 2.25~mag.

Fig.~\ref{fig_cmd} clearly shows the TRGB and the population of AGB stars immediately above it. To avoid the rich population of RGB stars, we select stars at least 0.1~mag brighter than the TRGB as measured in Fig.~\ref{fig_cmd}, which translates into
\beqa
\fH &<& 18.14~\mbox{\rm mag}  \,. \label{eq_magcut}
\eeqa

In addition, we adopt a blue color cut designed to limit contamination by young red supergiants (RSGs) and foreground dwarfs:
\beqa
\fJH &>& 0.88~\mbox{\rm mag} \label{eq_colcut}
\eeqa
We recall that the bulk of (essentially unreddened) foreground Milky Way stars are located in a vertical sequence at $\fJH=0.7$~mag \citep{Dalcanton_etal12, williams14}, which is also evident in Fig.~\ref{fig_cmd}.

Even with the above-mentioned color cut, a whole class of young luminous giants may still contaminate our list of AGB candidates. These include massive RSGs, and possibly also intermediate-mass red helium-burning stars, which are as bright as TP-AGB stars and which can become red enough (as a result of high metallicity or high reddening) to enter into our selection box \citep[see][]{Dalcanton_etal12}. However, those contaminants belonging to star clusters might be avoided by simply eliminating the youngest clusters from our discussion, as we show later in Sect.~\ref{results}. These selection criteria will also miss many of the so-called extreme-AGB stars, which are reddened and dimmed by several magnitudes in the NIR, owing to their own circumstellar dust shells. These stars will be fainter than the NIR TRGB, but will be discussed with their Spitzer IR photometry in a forthcoming paper (Goldman et al., in prep.).

We have explored similar criteria to select candidate AGB stars from the optical observations as well:
\beqa
\fI &<& 0.09828\,(\fBI)^2 \\ \nonumber
    & & -0.3578\,(\fBI) + 20.455 \\
\fBI &>& 21.675 - 0.91875\,\fI
\eeqa
However this sample is less complete than the NIR one, and includes a significant number of ``optical AGB candidates'' which are actually far away from the region selected in the NIR CMD of Fig.~\ref{fig_cmd}. A similar problem was recently uncovered by \citet{boyer19}, who shows how NIR-bright M stars move to magnitudes fainter than the TRGB in the F814W passband (see their figure 6). Therefore, we exclude the optical criteria from our discussion.

Based on the NIR selection, we identify 255\,671 candidate NIR AGB stars in the entire PHAT 6-filter catalogue, which covers nearly 1/3 of the M31 disk. $\sim\!150\,000$ of these stars belong to the bulge area (mainly to Brick~1). This alone is an impressive statistic, to be compared to the $\sim\!61\,000$ and $8\,500$ candidate TP-AGBs identified in the LMC and SMC, respectively, using similar NIR criteria \citep[][]{cioni06lmc,cioni06smc,boyer11}. The density of AGB candidate stars decreases nearly exponentially with galactocentric radius. In addition, there is a strong concentration in the bulge, and some mild concentrations along the 10-kpc ring and the overdensity at 3.5~kpc identified by \citet{davidge12}. 
The ratio of AGB stars to RGB stars varies with radius, and is twice as high in the outer disk as in the inner disk. This is probably caused by a combination of the larger metallicities (expected shorter TP-AGB lifetimes) and larger fraction of older populations (again shorter TP-AGB lifetimes) in the inner M31 disk. 

\subsection{Matching AGB stars and clusters}
\label{sec_match}

The 255\,671 AGB stars will be a mixture of numerous field AGB stars and a smaller number that are true cluster members. We identify this latter population by looking at all possible AGB--cluster matches and looking for excesses above what is expected for a pure field-AGB population.

\begin{figure}
\includegraphics[width=0.49\textwidth]{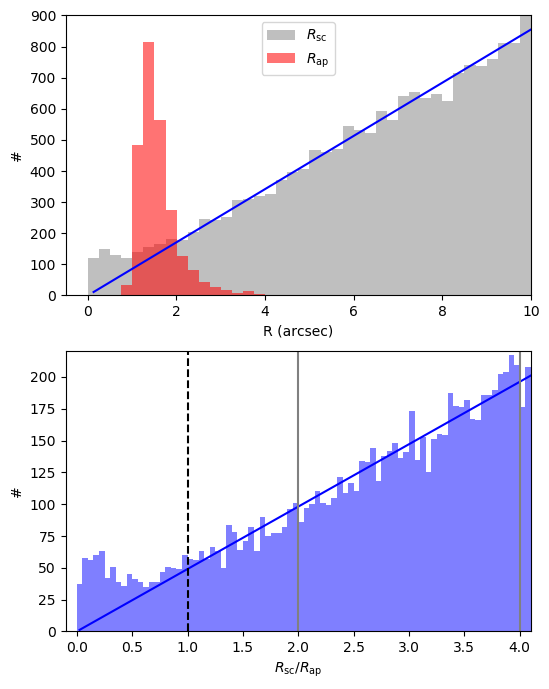}
\caption{{Top panel:} Distribution of separations between candidate AGB stars and cluster centres, \Rsc, for the 2491 candidates found with separations smaller than 10\arcsec\ (gray histogram). The plotted line shows the linear distribution expected from random matches; its mean slope is derived from all stars with $\Rsc$ between 6 and 60\arcsec. For comparison, the red histogram shows the distribution of cluster aperture radii \Rap\ in the AP catalog. {Bottom panel:} Distribution of the ratio between separation and the aperture radius, $\Rsc/\Rap$. The gray continuous vertical lines delimit the region used to estimate the mean field density, while the dashed line marks the upper limit adopted for building the present catalogue.
}
\label{fig_separation}
\end{figure}

We cross-match our list of candidate AGB stars with the AP cluster catalogue, initially keeping all candidates within a radius $\Rsc<60$\arcsec\ (225~pc) from the optically-determined cluster centres, which results in over 615\,000 AGB-cluster pairs. The top panel of Fig.~\ref{fig_separation} shows the histogram of \Rsc\ values of AGB-cluster pairs up to 10\arcsec, together with the distribution of aperture radii, \Rap, for the clusters. We find that for all $\Rsc\gtrsim4\arcsec$ the number of pairs increases linearly with $\Rsc$, as expected from a sample dominated by random matches between clusters and uniformly-distributed field stars. Fig.~\ref{fig_separation}, however, clearly shows that there is an excess of matches for $\Rsc<2$\arcsec. The bulk of clusters in the AP catalog has aperture radii, \Rap, ranging from 1\arcsec\ to 2\arcsec, as also shown in Fig.~\ref{fig_separation}. It becomes evident that there is an excess of AGB candidates observed at $\Rsc\la\Rap$, with respect to the numbers expected from a uniform distribution. By extending the line obtained for large $\Rsc$ towards the $\Rsc=0, n(\Rsc)=0$ limit (Fig.~\ref{fig_separation}, top panel), we can even determine that the excess of matches above this line amounts to 438 stars.

We repeat the exercise of measuring the excess, but using separations normalized by the size of the cluster, \Rap. The distribution of the $\Rsc/\Rap$ ratio is shown in the bottom panel of Fig.~\ref{fig_separation}. A total of 937 AGB candidates are found at $\Rsc/\Rap<1$. This time we decide to measure the field stellar density in an annulus extending from 2 to 4 times \Rap. Extrapolation of the field stellar density down to the $\Rsc/\Rap=0, n(\Rsc/\Rap)=0$ limit reveals that the excess of AGB stars starts at $\Rsc/\Rap$ values slightly below 1, although it becomes really evident only at $\Rsc/\Rap<0.6$. The excess amounts to $477\pm10$, which is about half of the number of TP-AGB stars observed at $\Rsc/\Rap<1$, and comparable to the number inferred from the analysis of \Rsc.

We therefore infer that the AP clusters contain about 450 likely AGB members inside their \Rap, and about the same number of ``field intruders''. That said, it is nearly impossible to say with certainty whether an individual AGB star in a cluster is a member or not. The approach to be followed must therefore be a statistical one, in which membership probabilities are assigned. In the discussion that follows, we adopt an inclusive criterion, considering all stars inside \Rap\ as possible cluster members. 

\begin{figure*}
\includegraphics[width=0.49\textwidth]{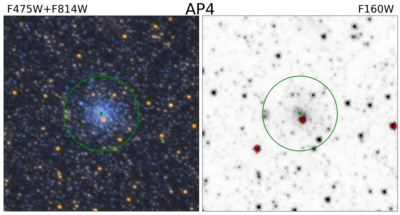} \hfill
\includegraphics[width=0.49\textwidth]{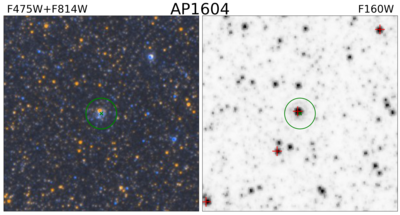} \\
\includegraphics[width=0.49\textwidth]{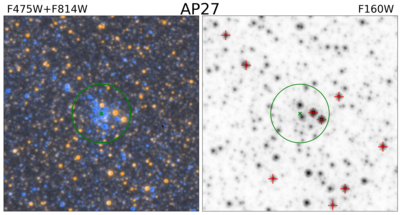} \hfill
\includegraphics[width=0.49\textwidth]{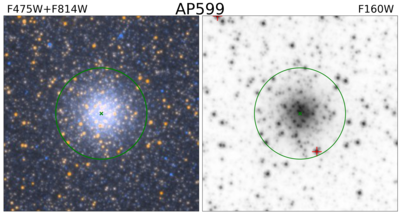}\\
\includegraphics[width=0.49\textwidth]{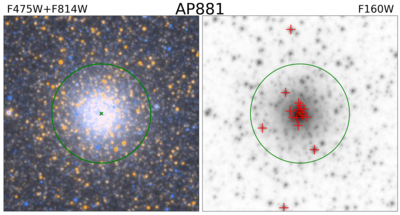} \hfill
\includegraphics[width=0.49\textwidth]{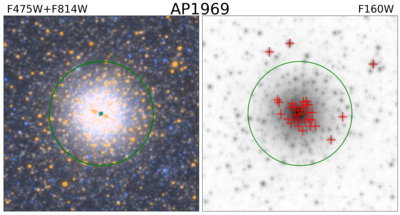} 
\caption{Examples of AP clusters containing AGB candidates. For each cluster, the left panel shows a false-color representation of the PHAT F475W+F814W optical images, which was used to find the clusters and derive their morphological parameters in \citet{johnson15}, while the right panel shows the F160W image, whose photometry is crucial for identifying the AGB stars in this work. Images are centered on the cluster and have dimensions of $15\arcsec\times15\arcsec$ (or $5.7\,\mathrm{pc}\times5.7\,\mathrm{pc}$), with North up and East left. In green we mark the cluster centre and aperture radius \Rap, in red (in the F160W image only) all the candidate AGB stars in the area. Six examples are shown: AP4 (the first entry in the catalogue) and AP1604 illustrates the almost-perfect superposition between intermediate-age clusters and lone, very bright and red AGB stars. We note the unusual compactness of AP1604, which makes the association particularly compelling. AP27 is a young cluster with two candidate AGB stars clearly standing out due to their NIR brightness; both are also the brightest red giants in the composite optical image. AP599, AP881 and AP1969 represent a sequence of globular-like clusters of increasing central density and likely old age (see \url{https://www.cfa.harvard.edu/oir/eg/m31clusters/phat}). AP599 (B201-G250) has a lone AGB candidate slightly off-centre, despite sitting in a field in which AGB stars are actually rare. AP881 (B218-G272) and AP1969 (B225-G280) are old globular clusters with many bright red giants, 14 and 29 of which, respectively, are candidate AGB stars. It is clear that their centremost candidates could be affected by crowding errors and even by blending of multiple NIR sources; however, the outermost sources inside \Rap\ are free from these problems, and likely cluster members.}
\label{fig_images}
\end{figure*}

\begin{figure*}
\includegraphics[width=0.49\textwidth]{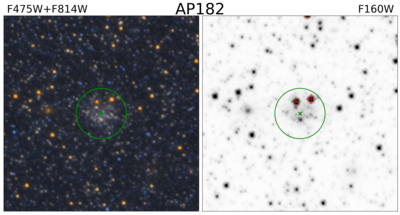} \hfill
\includegraphics[width=0.49\textwidth]{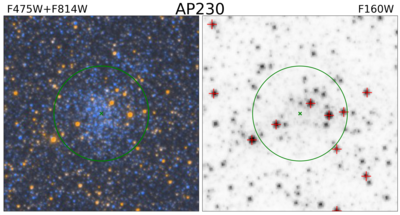} \\
\includegraphics[width=0.49\textwidth]{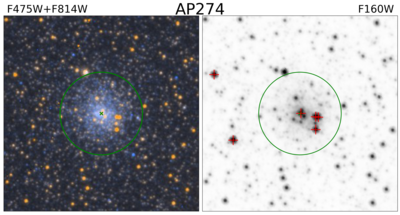} \hfill
\includegraphics[width=0.49\textwidth]{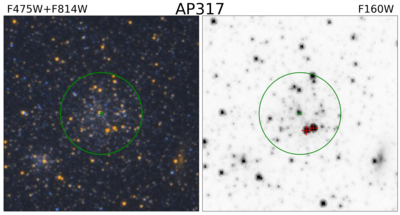}
\caption{Same as in Fig.~\ref{fig_images}, but now illustrating cases in which the association between stars and clusters is particularly compelling, thanks to the multiple AGB candidates on the cluster area, the uncrowded images, and the low field contamination.}
\label{fig_images_multiple}
\end{figure*}

\begin{figure*}
\includegraphics[width=0.49\textwidth]{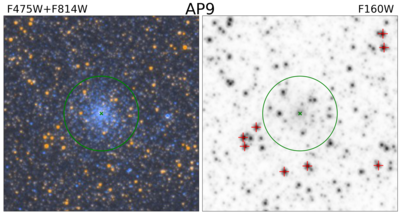} \hfill
\includegraphics[width=0.49\textwidth]{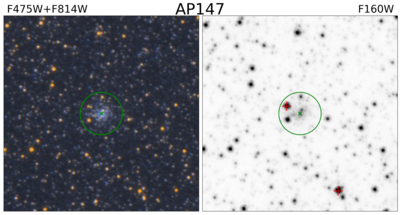} \\
\includegraphics[width=0.49\textwidth]{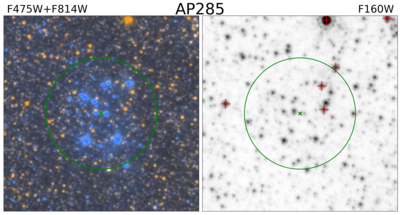} \hfill
\includegraphics[width=0.49\textwidth]{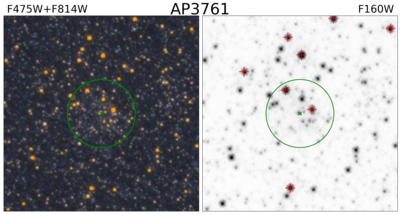}
\caption{Same as in Fig.~\ref{fig_images}, but now illustrating negative and potentially-problematic results: First, we show AP9 as an example of non-detection: this cluster is well populated and has the right age to contain AGB stars ($\log t/\mathrm{yr}$ between 8.6 and 9.1), but none is close enough to classify as a likely cluster member. AP147 contains one candidate AGB with F160W$=18.111$~mag, which is just slightly above the F160W limit we define. However, the optical images (which have a better resolution than the F160W one) indicate that the NIR source might be affected by the near-blending of two red giants. Although its \fcrowd\ parameter is of just 0.048~mag, this introduces enough uncertainty in its NIR magnitude that this star could actually be fainter than the F160W magnitude limit we set for the AGB sample. AP285 instead contains 2 bright NIR sources inside the cluster \Rap\, but it is probably too young ($\log t/\mathrm{yr}$ between 6.45 and 6.75) to contain genuine AGB stars. These are more likely reddened RSGs, or field AGB stars falling inside the large \Rap\ of this cluster. Finally, AP3761 has two candidate AGB stars but suffers from a more generic problem: clusters with high AP numbers tend to be fuzzier, less massive, and to have badly-defined limits, compared to those with low AP numbers. These uncertainties affect the estimates of the cluster center and \Rap, and hence the membership probabilities, ages and mass estimates.}
\label{fig_images_no}
\end{figure*}

Figures~\ref{fig_images} and \ref{fig_images_multiple} illustrate a few ``good cases'' in which the association between candidate AGB stars and clusters appears evident, especially in the latter, for which candidates are multiple and could hardly result from field contamination or crowding. Fig.~\ref{fig_images_no} instead shows an example in which no AGB star was found inside \Rap\ despite the favourable conditions, plus a few somewhat dubious cases of association. There are many other situations in which the association between the AGB stars and the clusters -- or even the AGB classification -- could be questioned using either photometric or astrophysical arguments. In the following, we will discuss a few aspects that can help us to distinguish between likely and unlikely associations.

Before proceeding, however, we note that both the cluster centres (used to derive \Rsc) and \Rap\ were measured from the optical images, and hence they are unlikely to be strongly affected by the presence of some very few NIR-bright candidate AGB stars in their neighbourhood. Indeed, this appears clear from the images in Figs.~\ref{fig_images} to \ref{fig_images_no}.

\subsection{Characterization of crowding}
\label{sec_crowd}

An excess of bright NIR stars inside the cluster cores could well be caused by crowding (understood as photometric errors plus stellar blends), rather than by the presence of bright cluster members. Globular clusters in the Milky Way provide good examples of crowding affecting the brightness of red giants in infrared passbands \citep[see, e.g.,][]{boyer10}. Fortunately, the work by \citet{williams14} allows us to make a first evaluation of these effects on our catalogue.

\begin{figure}
\includegraphics[width=0.49\textwidth]{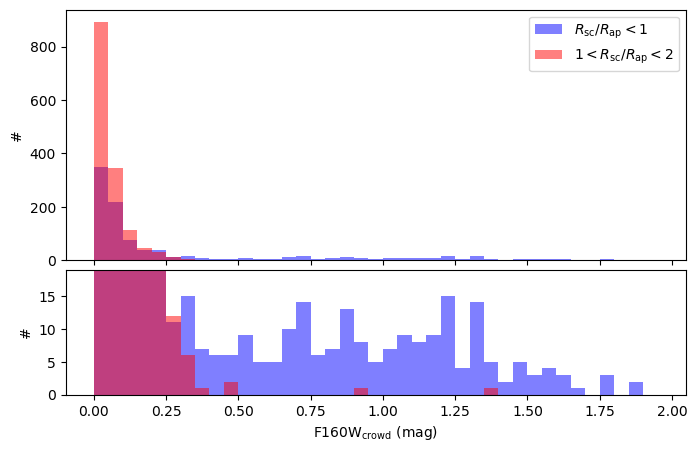}
\caption{The top panel presents a histogram of the parameter \fcrowd\ for candidate AGB stars both inside (blue) and immediately outside \Rap\ (red). The bulk of the distribution is located at $\fcrowd<0.2$~mag, which overalls points to a modest impact of crowding in the NIR photometry of the entire sample. The bottom panel zooms on the tail of large \fcrowd\ values presented by the sample inside \Rap.}
\label{fig_crowding}
\end{figure}

The six filter photometry catalog from \citet{williams14} has a crowding parameter in each filter that characterizes how much brighter the star would have been had nearby stars not been fit simultaneously. For the F160W filter, this parameter is the \verb$F160W_crowd$ parameter in the data bases, which we refer to here as F160W$_{\rm crowd}$. Fig.~\ref{fig_crowding} shows the distribution of this parameter for the samples of candidate AGB stars both inside and outside clusters. Stars outside clusters ($\Rsc>\Rap$) have \fcrowd\ typically constrained below 0.2 mag. Stars inside clusters present nearly the same distribution of crowding parameters but have an additional long tail of higher \fcrowd\ values, with 260 stars (27~\% of the sample) having $\fcrowd>0.2$~mag. This behaviour is closely followed by F110W$_{\rm crowd}$. This indicates that in-cluster crowding is probably affecting a good fraction of our sample. We have flagged in our catalogs the objects (265 out of 937) for which the sum of F160W and \fcrowd\ could put stars below the minimum magnitude for being classified as an AGB, that is $\mathrm{F160W}+\fcrowd>18.1$~mag. One such example is the AGB star in AP147, illustrated in Fig.~\ref{fig_images_no}. In a similar way, we also flag 33 additional stars for which the ``uncrowded color'' would bee too blue, i.e., $\mathrm{F110W}+\mathrm{F110W}_{\rm crowd}-\mathrm{F160W}-\fcrowd<0.88$~mag. This is not a determinant criterion to exclude stars from our catalogue, however, since the photometry was correctly done using all neighbouring sources. This is just a flag to pinpoint objects where the photometry and AGB classification are \textit{possibly} more affected by crowding, than usual.

\begin{figure}
\includegraphics[width=0.49\textwidth]{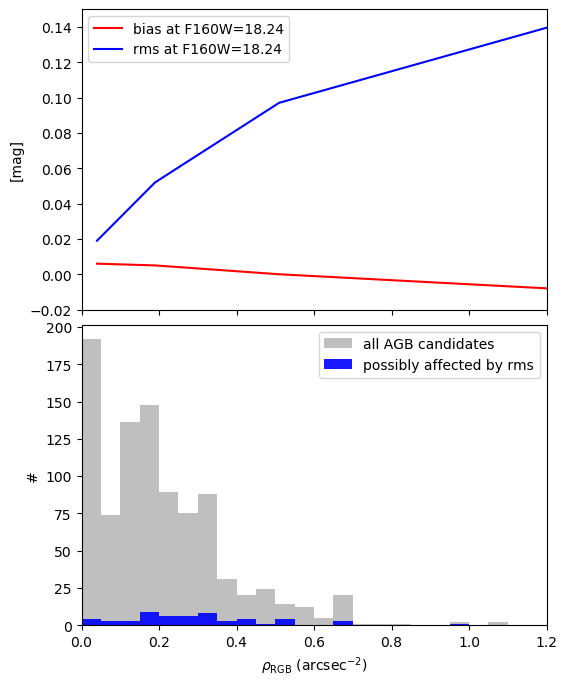}
\caption{Top panel: the magnitude bias and rms error for stars at the TRGB ($\mathrm{F160W}=18.24$~mag), plotted over the interval of RGB densities relevant for the star clusters. The curves are determined by linear interpolation of the values determined by \citet{williams14} using artificial star tests of $\mathrm{F160W}=18.0$~mag and $\mathrm{F160W}=18.5$~mag).
Bottom panel: the gray histogram shows the distribution of RGB densities for all 937 candidate AGB stars in clusters. The blue histogram shows the 55 stars whose magnitudes are compatible with a TRGB star being scattered upwards by the same amount as the rms. Similarly, we checked which AGB candidates could be stars at the TRGB moved into the AGB region by the photometric bias, but no such case was found.}
\label{fig_bias}
\end{figure}

To better characterize the level of crowding star-by-star, we can also employ the artificial stars tests performed by \citet[][see their section 5]{williams14} on PHAT stacked images. They were performed for a few representative values of ``RGB star densities'', or $\rho_\mathrm{RGB}$, intended as the number of stars of $18.5<\mathrm{F160W}<19.5$ per arcsec$^2$. Particularly relevant in the context of this paper are the ``magnitude random errors and biases'' for stars located very close to the TRGB, namely at $\mathrm{F160W}=18.24$~mag, which are plotted in the top panel of Fig.~\ref{fig_bias}. A magnitude bias $<-0.1$~mag could easily cause the numerous RGB stars present in the cluster and field to enter into the AGB area of the CMD. Similarly, a magnitude random error exceeding 0.1~mag would cause a sizeable fraction of these RGB stars to be randomly scattered in the same region. 

The bottom panel of Fig.~\ref{fig_bias} presents the distribution of stellar density of RGB stars with $18.5<\mathrm{F160W}<19.5$, $\rho_\mathrm{RGB}$, for our candidate AGB stars in clusters, measured inside the \Rap\ of each cluster. Many clusters have a so scarcely populated RGB that this parameter turns out to be null, which essentially points to negligible crowding in the cluster core compared to the environment of typical field AGB stars. More populous clusters have a significant number of RGB stars, but their $\rho_\mathrm{RGB}$ never exceeds 1.1~$\mathrm{arcsec}^{-2}$. At these maximum RGB densities, the magnitude bias turns out to be tiny (about $\sim-0.006$~mag), although the magnitude rms error (bottom panel of Fig.~\ref{fig_bias}) amounts to a very significant $\sim0.12$~mag. Assuming a Gaussian distribution of magnitude errors, the faintest among our candidate AGB stars could actually be TRGB stars (or real AGB stars slightly fainter than the TRGB) which entered into the AGB selection region just because of the photometric scatter. However, we verified that this fraction is very small: just 55 AGB candidates (shown in Fig.~\ref{fig_bias}), or $\sim6$~\% of the sample, have a F160W magnitude compatible with it being a TRGB star scattered upwards by one rms. These cases include AGB candidates in dense globular clusters such as AP1969 (B225-G280; Fig.~\ref{fig_images}).

Overall, these estimates of the cluster densities and crowding errors are quite crude and subject to many uncertainties: few RGB stars per cluster, no real fitting of cluster density profile, no real estimate of the likelihood of RGB stars being scattered into the AGB region, etc. Moreover, it suffices to examine Fig.~\ref{fig_images} to conclude that proper estimates of the effect of crowding would require extensive -- and time-consuming -- experiments of artificial star tests using the actual cluster images, or fully synthetic cluster tests. For the moment, we limit ourselves to making our $\rho_\mathrm{RGB}$ estimates available in the final catalog, so that potential users can appreciate, with the help of Fig.~\ref{fig_bias}, the chances that AGB candidates in a particular cluster are significantly affected by crowding.

\subsection{C/M classification with WFC3/IR medium-band filters}
\label{sec_CM}

We augment our catalog of cluster AGB candidates with estimates of their spectral subtypes where available. 

First, we use results from \citet{boyer13,boyer19}, who performed an HST survey using WFC3/IR medium-band filters in 21 regions across the PHAT footprint, to classify their bright red giants as either C- or M-type. Their method uses the sensitivity of the filters F127M, F139M and F153M to the presence of either CN+C$_2$ or H$_2$O absorption features in the near-infrared. We cross-match our sample of 937 candidate AGB stars in clusters with the complete \citet{boyer19} catalog. We find a total of 59 robust matches, for which the differences between the catalogs are smaller than 0.04\arcsec\ in position, and smaller than 0.017~mag in both F110W and F160W magnitudes. All of them are classified as M stars. The lack of C stars is this small sample is not surprising, given the remarkable scarcity of C stars in M31, especially in its innermost regions \citep{boyer19}. Indeed the average C/M ratio across the M31 disk amounts to just $\sim0.02$. 

\subsection{Keck/DEIMOS spectroscopy}
\label{sec_spectra}

\begin{figure*}
\includegraphics[width=0.49\textwidth]{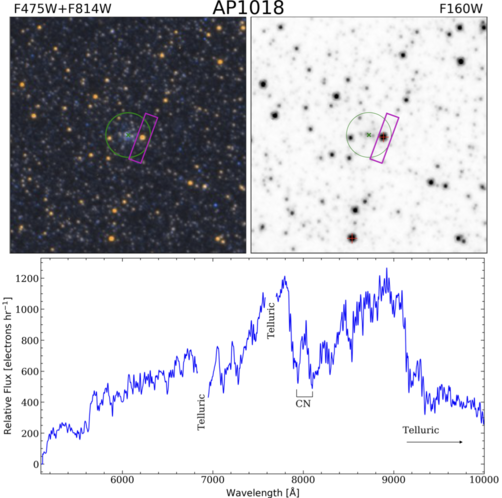} \hfill
\includegraphics[width=0.49\textwidth]{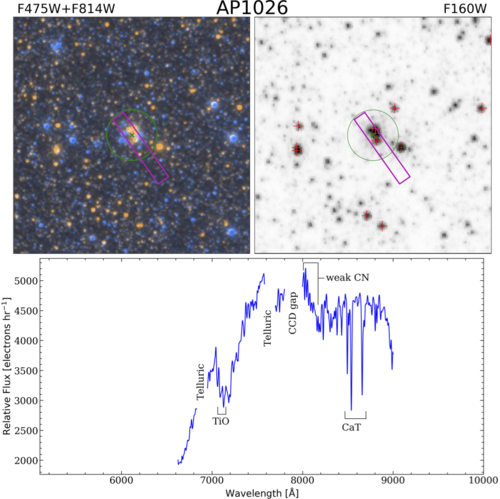} \\
\includegraphics[width=0.49\textwidth]{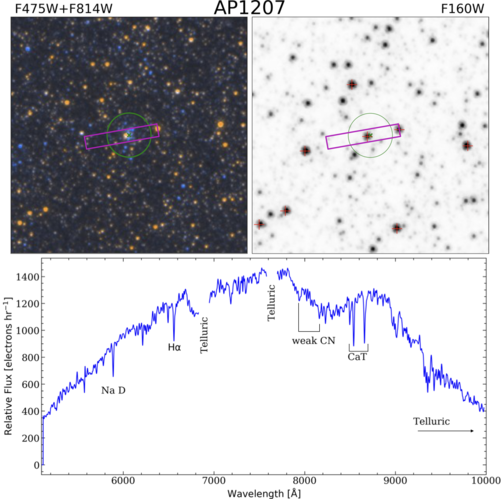} \hfill
\includegraphics[width=0.49\textwidth]{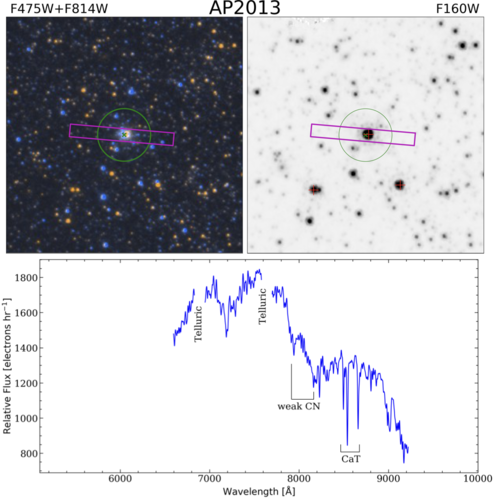} 
\caption{Examples of clusters for which we have spectra from Keck/DEIMOS. To the cluster images and AGB candidate identification in the top panels (as in Fig.~\ref{fig_images}), we add the position of the DEIMOS slit, and the extracted spectrum in the bottom panels. The most prominent spectral features are marked, and the main telluric features are either marked or removed.}
\label{fig_images_spectra}
\end{figure*}

As an extension of the Spectroscopic and Photometric Landscape of Andromeda's Stellar Halo (SPLASH) survey \citep[e.g.][]{raja05, raja06, gilbert06}, we used the Keck II 10-m telescope and DEIMOS spectrograph to obtain spectra of PHAT-selected star clusters and individual stars in the disk of M31 \citep{dorman12, dorman13, dorman15, quirk19}. The spectra fall into two categories: (1) low resolution spectra obtained with the 600ZD grating that cover the approximate wavelength range 4500--9500~\AA, and (2) medium resolution spectra obtained with the 1200G grating that cover the approximate wavelength range 6500--9000~\AA. In the entire database, we find 24 unique spectroscopic matches with clusters hosting AGB candidates. Some examples of spectra are shown in Fig.~\ref{fig_images_spectra} along with the corresponding star cluster image and DEIMOS slit overlay. It is important to note that the FWHM of seeing during the spectroscopic observations was in the range 0.5\arcsec--0.9\arcsec (for reference, the width of the DEIMOS slit is 0.8\arcsec) so these spectra are generally composite spectra of more than one bright cluster star. The TP-AGB star in the cluster AP1018 turns out to be a carbon star; note the strong ``W''-shaped CN spectral absorption feature at 7800--8200~\AA. The remaining three spectra shown display a much weaker version of this same ``W''--shaped CN spectral absorption feature, a feature that appears to be associated with evolved massive (5--10 \Msun) stars (Guhathakurta et al., in prep.).

Clear signatures of carbon stars are found only in the clusters AP1018 (see Fig.~\ref{fig_images_spectra}) and AP1508. These are low-mass clusters with very uncertain integrated-light ages, with 16 to 84~\% confidence intervals for $\log(t/\mathrm{yr})$ being 8.9--9.8 and 8.1--9.5, respectively. These two cases add to the very limited list of C stars in metal-rich open clusters, presently consisting of just four candidates in the Milky Way open clusters NGC\,2477 \citep{catchpole73}, NGC\,2660 \citep{hartwick73}, NGC\,7789 \citep{gaustad71} and Trumpler~5 \citep{kalinowski74}. 

\subsection{Variability}
\label{sec_variables}

The $160\arcsec\times160\arcsec$ field of view of the WFC3/IR F110W and F160W images is the smallest of the cameras/modes used in the PHAT project, and this sets the step size of the brick-based tiling used in the survey. By contrast, the field of view of the ACS/WFC F475W and F814W images is $205\arcsec\times205\arcsec$ which leads to significant overlap between adjacent pointings in these two filters. We have used time-resolved PHAT photometry (derived from constrained PSF fits to individual ACS exposures) in both F814W and F475W filters to search for variability among the AGB candidates that are located within star clusters in the overlap areas. For about 20~\% of them, the time baseline over which repeat F475W and F814W photometry is available is $\ga100$ days, which is comparable to the time scale on which AGB stars are expected to vary \citep{soraisam20}. The photometry time baseline for the remaining 80~\% of the stars is $<30$ days, which is short enough that we do not expect to detect significant variability. These two subsamples of AGB stars are hereafter referred to as the long time baseline (LTB) and short time baseline (STB) subsamples, respectively.

\begin{figure*}
\includegraphics[width=\textwidth]{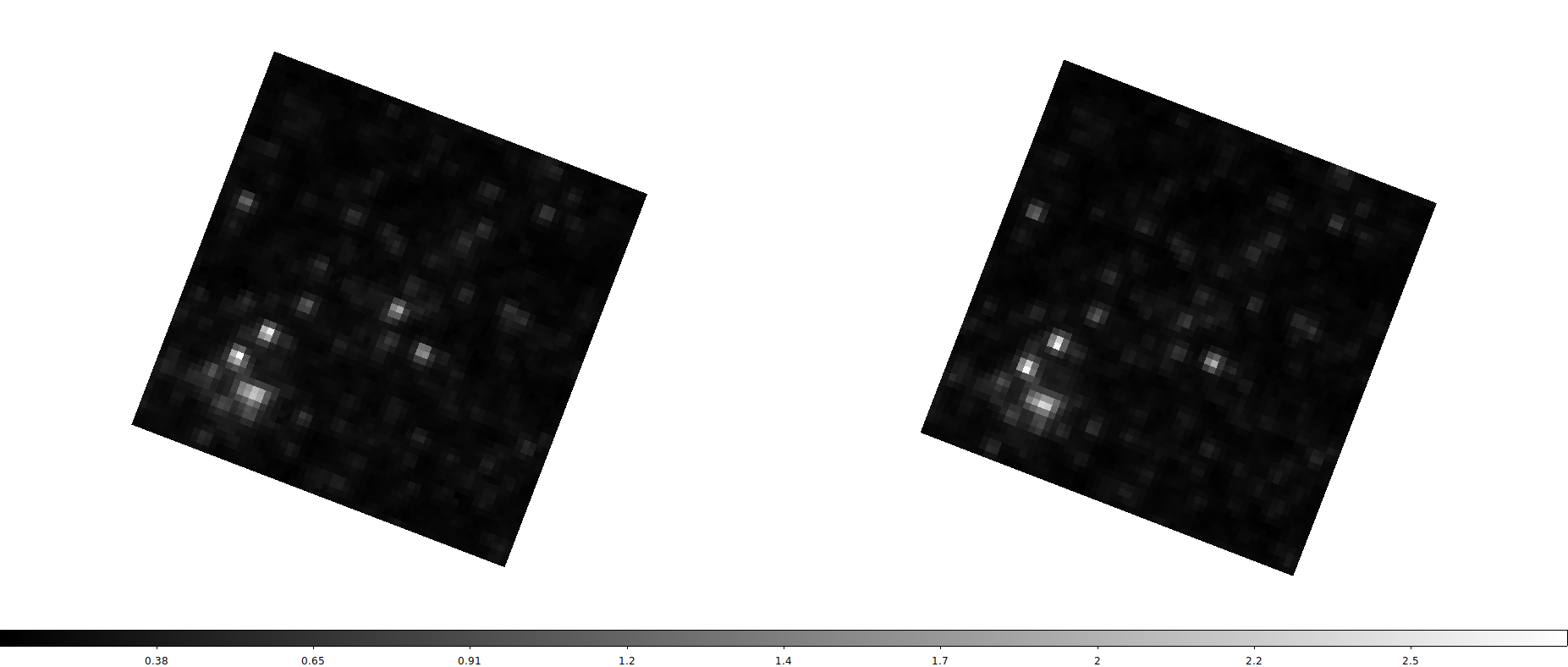} 
\caption{Two images of the AP2725 cluster in the ACS/WFC F475W filter, taken 201 days apart. The images are centred on the star PHAT10.9645575+41.500547, which is a candidate AGB star, and whose variation in brightness is evident.}
\label{fig_images_variable}
\end{figure*}

We use a chi-squared statistic as a measure of variability compared to the mean magnitude, and compute its value using the photometric measurements for each of our AGB stars in both F814W and F475W filters. We also characterize an average systematic error (e.g., due to crowding or cosmic ray hits) in the photometric measurements using the STB AGB stars in the sample. We use the 95th percentile of the distribution of chi-squared values of the STB AGBs in both F814W and F475W filters as thresholds for variability. We find 20 LTB AGBs above these thresholds. To confirm that this variability is not affected by the presence of some other bright star/variable star nearby or the crowding of the cluster, we look at individual HST images (in both F814W and F475W). Of the 20 stars, we find 10 of them as secure variables. One evident example is presented in Fig.~\ref{fig_images_variable}. The other 10 stars are affected by the crowding of the clusters or bad quality exposures. The details of this analysis along with a population of all detected variables in PHAT clusters (including the non-AGB ones) will be published in Mukherjee et al.\ (in prep).

\section{Sample properties}
\label{results}
 
AGB cluster candidates are only expected to be real AGB stars for cluster ages larger than $\sim100$~Myr. In younger clusters, post-He-burning stars do not develop a carbon-oxygen (or oxygen-neon in the case of the super-AGB stars) electron-degenerate core \citep{herwig05}, and as a consequence have a much quicker evolution up to pre-supernova stages. Moreover, across the entire age range in which AGB stars appear, their numbers are expected to vary as a consequence of multiple processes (third dredge-up, hot-bottom burning, carbon-star formation, onset of fundamental-mode high amplitude pulsation, dust-driven stellar winds, etc.) which occur with varying efficiency as a function of stellar mass and metallicity during the thermally-pulsing phase \citep[see e.g.,][]{marigo07}. Revealing this sequence of AGB star frequency and properties as a function of stellar age (and mass), and possibly also of metallicity, is actually one of the best reasons to define samples of AGB stars in clusters, as in the present paper. 

In this section, we attempt for a first interpretation of the data in terms of cluster ages, $t$. For a given age, the numbers of AGB stars observed in clusters should also scale with the cluster total mass, $M_\mathrm{tot}$, so we need to characterize both quantities before dealing with the AGB stars themselves.

\subsection{Age and mass distributions of clusters} 
\label{sec_clusterages}
 
We take the quantities $t$ and $M_\mathrm{tot}$ derived by \citet{Fouesneau_etal14} and \citet{beerman15} for the clusters in the AP catalog. More specifically, we use the pair of values that corresponds to their ``best method'', which derive either from fitting of isochrones in the case of clusters with a good-quality CMD, or from the the integrated optical magnitudes and colors when the CMD is deemed too uncertain. Out of the 2753 clusters in our initial catalog, 2736 have these quantities determined. CMD ages \citep[from][]{johnson16} are available for 1253 clusters, generally corresponding to the youngest part of the cluster sample.

\begin{figure*}
\includegraphics[width=0.49\textwidth]{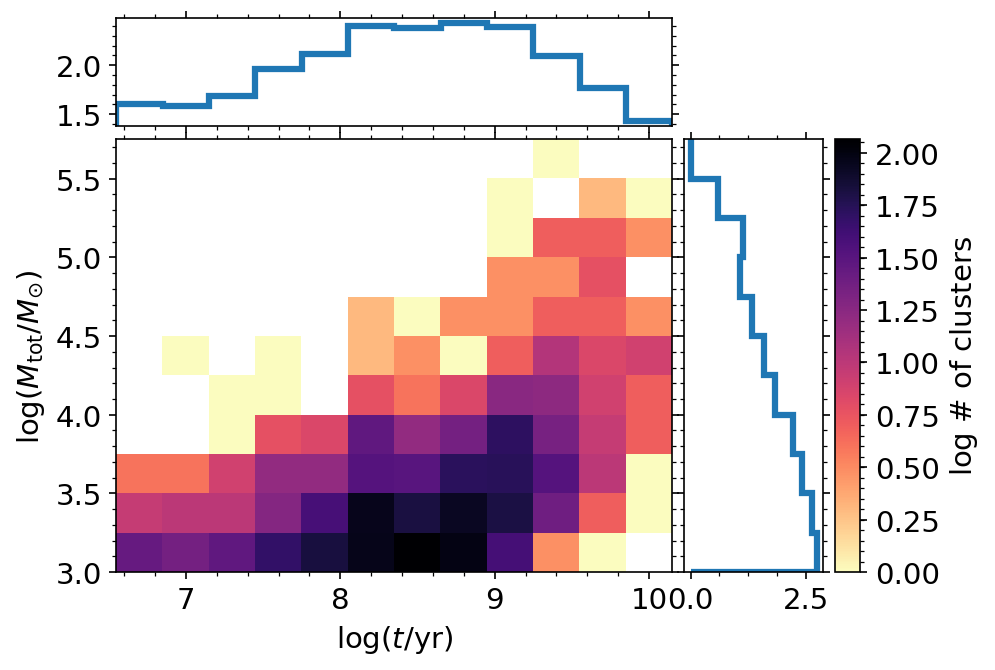} \hfill
\includegraphics[width=0.49\textwidth]{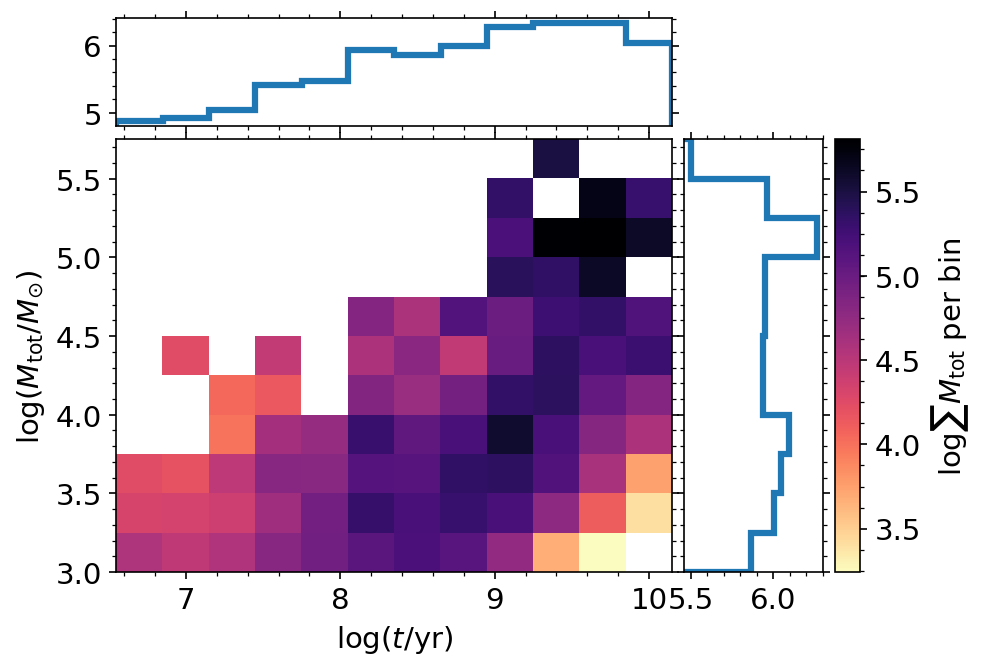} \\
\includegraphics[width=0.49\textwidth]{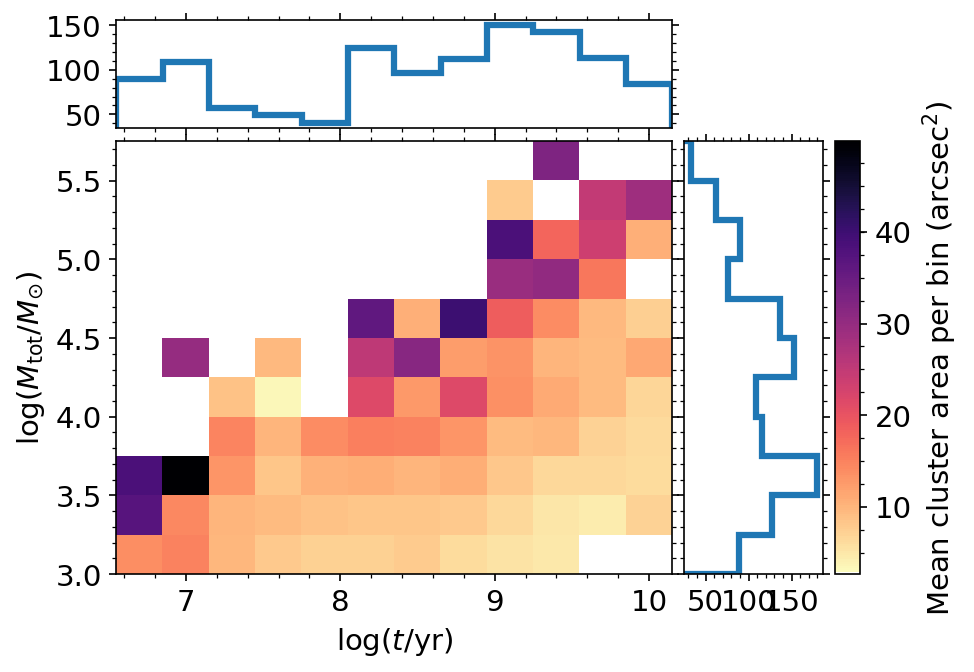} \hfill
\includegraphics[width=0.49\textwidth]{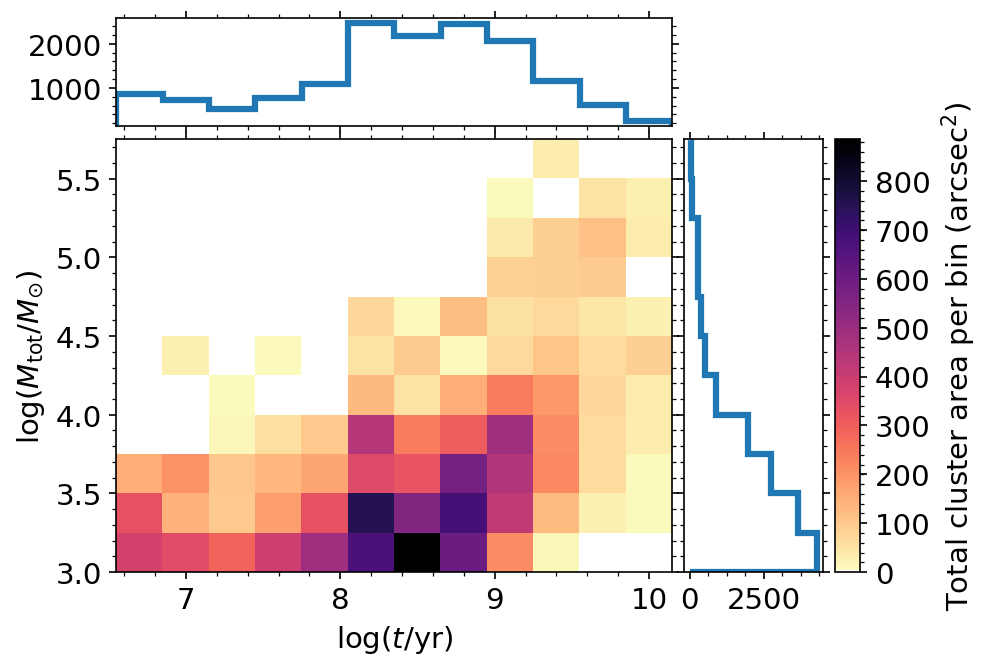}
\caption{Distribution of the AP cluster data used in this work (not in bulge, with NIR photometry, with ages) on the  $\log t$ versus  $\log M_\mathrm{tot}$ plane. They include all 2736 clusters available, regardless of the presence of AGB candidates. The top left panel shows the number of clusters per age-mass bin and histograms for both mass and age. The top-right panel shows the same for the total mass comprised in every bin. The bottom-left panel shows the mean area inside $\Rap$, for all clusters in a bin. The bottom-right panel shows the total area inside $\Rap$ for all clusters in a bin.
}
\label{fig_2dclusters}
\end{figure*}

Figure~\ref{fig_2dclusters} shows the global properties of the clusters used in this work, in the $\log t$ versus $\log M_\mathrm{tot}$ plane. As can be noticed, the sample includes large numbers of young and intermediate-age low-mass clusters (with $M_\mathrm{tot}\la10^4$~\Msun, see top left panel), but the total mass of the cluster sample is instead concentrated at ages older than a few Gyr (see top right panel), which include clusters with masses in excess of $M_\mathrm{tot}\ga10^5$~\Msun. There are also clear trends in the mean sky area, $\pi\Rap^2$, covered by clusters of different age and mass, as shown in the bottom-left panel. Clusters with larger $M_\mathrm{tot}$ tend to cover larger areas, but this also happens for very young clusters, with ages $t\la10^7$~yr. Together with the distribution of numbers of cluster (top-left panel), these trends give origin to the distribution of total area depicted in the bottom-right panel. 

Whenever necessary, we use the estimated $1\sigma$ errors in age and mass as being half the difference between the 16\% and 84\% percentiles in their probability density functions in logarithmic space. We recall that the age determinations coming from integrated photometry, by their own nature, are uncertain, with typical errors amounting to $\sim0.2$~dex. But again, this situation compares well with the situation for  the Magellanic Cloud clusters in the 1980s, for which the best ages and mass estimates were coming from integrated photometry rather than from their (very noisy, at the time) CMDs.

Before proceeding, we should define the most suitable age binning for the clusters in our analysis. In Fig.~\ref{fig_2dclusters}, and in the following, we choose a bin width of 0.3~dex in $\log t$ (or a factor of 2 in linear age) primarily because narrower bins would decrease the numbers of AGB candidates to values below $\sim10$ at some age intervals, producing significantly noisier histograms for the quantities we are going to measure. $\sim80$~\% of the clusters with age determinations have $1\sigma$ errors smaller than this bin width. 

\subsection{Age distribution of AGB candidates} 
\label{sec_agbages}
  
\begin{figure*}
\includegraphics[width=0.49\textwidth]{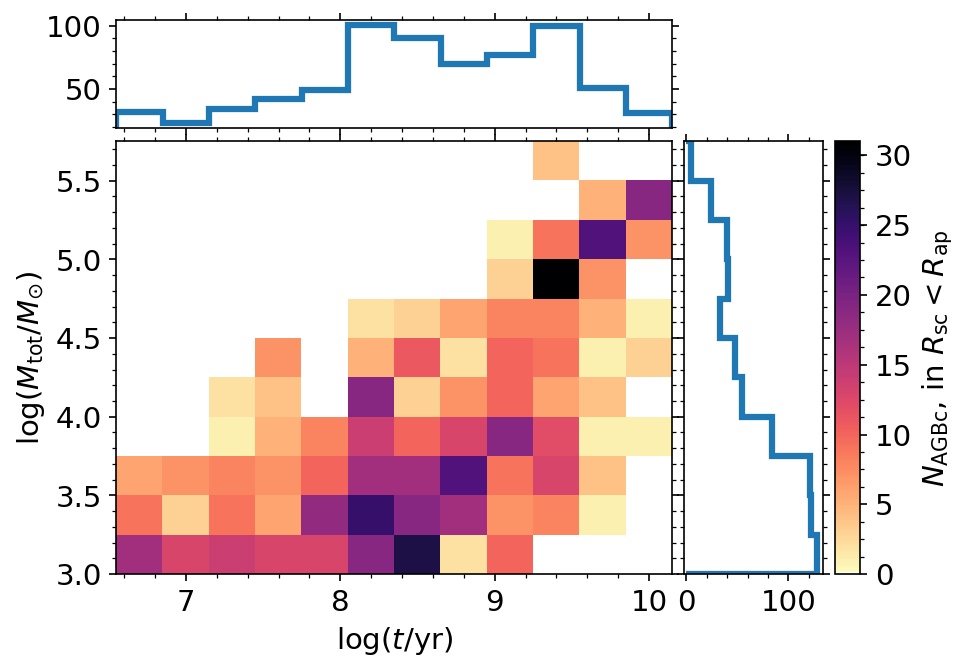} \hfill
\includegraphics[width=0.49\textwidth]{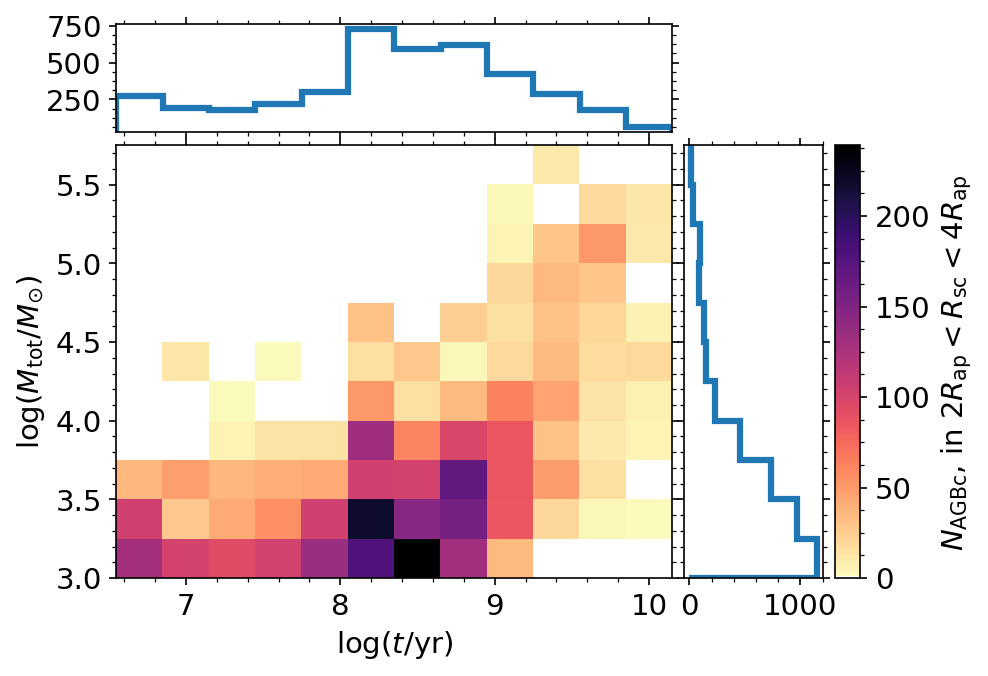} \\
\includegraphics[width=0.49\textwidth]{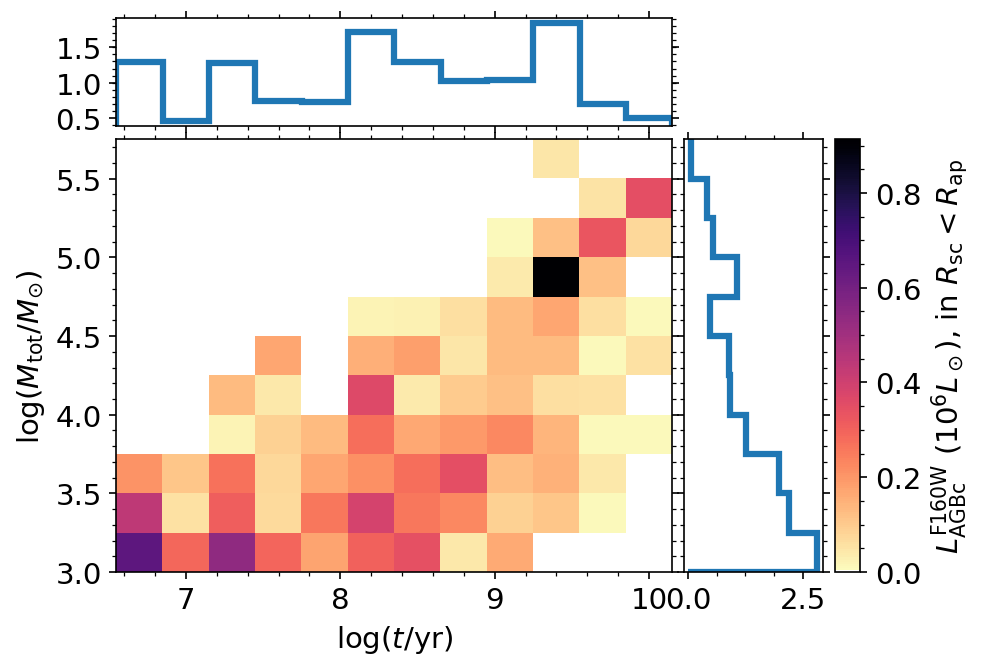} \hfill
\includegraphics[width=0.49\textwidth]{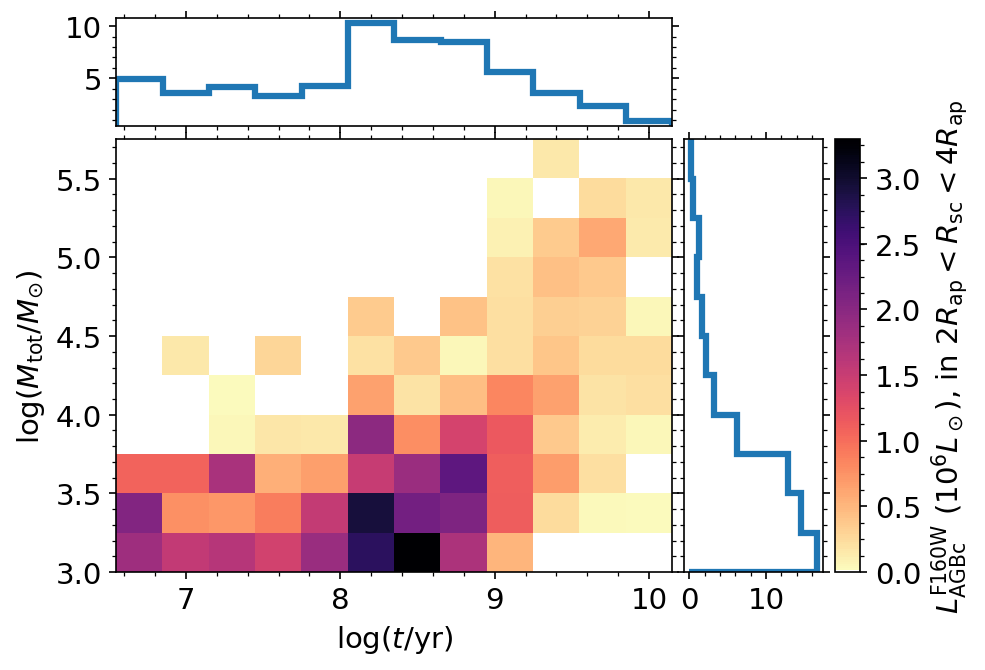}
\caption{Top row: Distribution of the AGB candidates in the $\log t$ versus $\log M_\mathrm{tot}$ plane, together with cumulative histograms for both quantities. The top left panel shows the number of AGB candidates inside the $\Rsc<\Rap$, while the top-right shows the numbers in the ``background sample'' at $2\Rap<\Rsc<4\Rap$, for which we attribute the same cluster ages and masses as for the stars in $\Rsc<\Rap$. The bottom row shows the same for the total luminosity of these stellar samples in the F160W filter.
}
\label{fig_2dAGBc}
\end{figure*}

AGB candidates inside $\Rap$ are attributed the same age $t$ as their hosting cluster, and associated to a stellar population parent mass equal to $M_\mathrm{tot}$. The top-left panel of Fig.~\ref{fig_2dAGBc} shows the distribution of their numbers in the $t$ versus $M_\mathrm{tot}$ plane. This distribution somewhat reflects the distribution of cluster total masses already shown in Fig.~\ref{fig_2dclusters}. More remarkable is the fact that AGB candidates are concentrated in the age interval between $\log(t/\mathrm{yr})=8.1$ and 9.3.
Comparison with the numbers of clusters in Fig.~\ref{fig_2dclusters} reveals that the mean numbers of AGB candidates per cluster varies between 0 and 0.7.

To proceed in our considerations, we also need to define a ``background model'', made of stars likely not associated with the clusters, but which are treated in the same way as the stars (real members or not) located in the cluster centres. We define this background model using the AGB candidates located in an annulus with $2\Rap<\Rsc<4\Rap$. This annulus is far enough from the cluster centre to be dominated by field stars (see Fig.~\ref{fig_separation}), and small enough to not reflect the strong variations in the field density and mean extinction of star-forming regions in the M31 disk. 

5884 candidate AGB stars are found in the $2\Rap<\Rsc<4\Rap$ area around 599 clusters. This sample turns out to present an age distribution clearly different from that of stars inside $\Rap$, as shown in the top-right panel of Fig.~\ref{fig_2dAGBc}. And, not surprisingly, their age-$M_\mathrm{tot}$ distribution resembles very much the distribution of cluster areas depicted in the bottom-right panel of Fig.~\ref{fig_2dclusters}. This indicates that they are indeed sampling the background population of field AGB candidates, in proportion to the $12\pi\Rap^2$ area around every cluster. 

We proceed computing the total F160W luminosity of the candidate AGB stars:
\begin{equation}
    L_\mathrm{AGBc}^\mathrm{F160W} = \sum_i 10^{-0.4\,(\mathrm{F160W}_i-24.50-3.37)} 
\end{equation}
where 24.50~mag is the adopted distance modulus and 3.37~mag is the solar absolute magnitude in F160W \citep{willmer18}. This quantity is computed for both the in-cluster ($\Rsc<\Rap$) and the background ($2\Rap<\Rsc<4\Rap$) samples, and shown in the bottom row of Fig.~\ref{fig_2dAGBc}.

We finally derive two quantities of direct interest for studies of stellar populations: the numbers of AGB stars and their total luminosities per unit mass of the searched clusters, corrected by the contamination from field stars, for every age bin:
\begin{eqnarray}
\frac{N_\mathrm{AGBc}}{M_\mathrm{tot}} &=& 
    \frac
    {\sum_i \left[N_\mathrm{AGBc,in} - (A_\mathrm{in}/A_\mathrm{out}) N_\mathrm{AGBc,out}\right]}
    {\sum_i M_\mathrm{tot}} \label{eq_n}  \,\,, \label{eq_nagb} \\
\frac{L_\mathrm{AGBc}^\mathrm{F160W}}{M_\mathrm{tot}} &=& 
    \frac
    {\sum_i \left[L_\mathrm{AGBc,in}^\mathrm{F160W} - (A_\mathrm{in}/A_\mathrm{out}) L_\mathrm{AGBc,out}^\mathrm{F160W}\right]}
    {\sum_i M_\mathrm{tot}} \label{eq_lum} \,\,, \label{eq_lagb}
\end{eqnarray}
where `in' stands for the $\Rsc<\Rap$ region, `out' for the $2\Rap<\Rsc<4\Rap$ annulus, and $A_\mathrm{in}/A_\mathrm{out}=1/12$ for their relative areas. The summations are done over all clusters $i$ inside every age bin, including those without any AGB candidate in their proximity.

\begin{figure}
\includegraphics[width=0.49\textwidth]{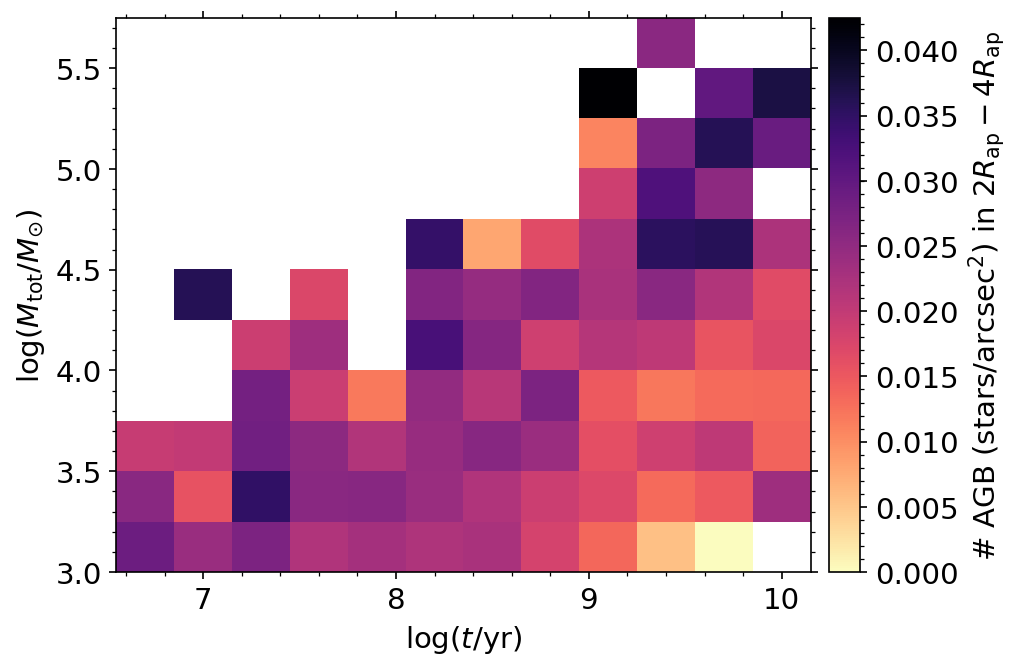} \\
\includegraphics[width=0.49\textwidth]{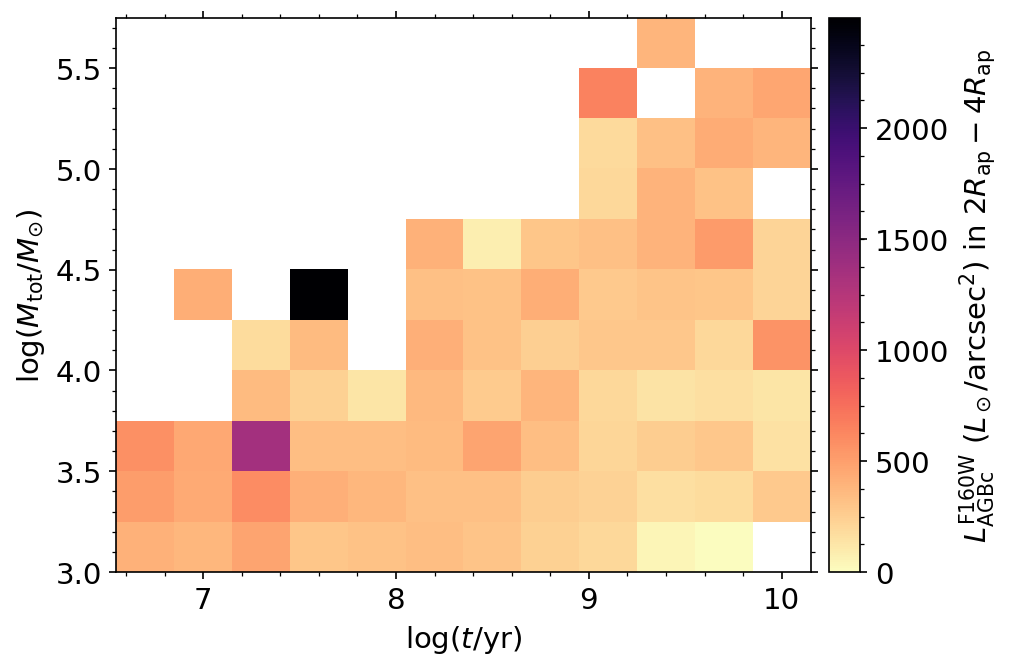}
\caption{The densities of background field stars contributing to Eqs.~\ref{eq_n} and \ref{eq_lum}. The top panel is for the number of AGB candidates, the bottom panel is for their total luminosities in the F160W filter.
}
\label{fig_2ddensities}
\end{figure}

All quantities needed to compute Eqs.~\ref{eq_n} and \ref{eq_lum} are illustrated in the previous Figs.~\ref{fig_2dclusters} and \ref{fig_2dAGBc}. In particular, $N_\mathrm{AGBc,in}$ and $L_\mathrm{AGBc,in}^\mathrm{F160W}$ are illustrated in the left panels of Fig.~\ref{fig_2dAGBc}, while $M_\mathrm{tot}$ is in the top-right panel of Fig.~\ref{fig_2dclusters}.
The bulk contribution of the background stars to Eqs.~\ref{eq_n} and \ref{eq_lum} (i.e., the terms $(A_\mathrm{in}/A_\mathrm{out}) N_\mathrm{AGBc,out}$ and $(A_\mathrm{in}/A_\mathrm{out})L_\mathrm{AGBc,out}^\mathrm{F160W}$) in every mass and age bin, can be derived by simply multiplying the numbers in the right panels of Fig.~\ref{fig_2dAGBc} by (1/12). In addition, we illustrate in Fig.~\ref{fig_2ddensities} the densities of star counts and of the F160W luminosities in the background sample. These densities differ from an uniform distribution because of the Poisson noise, and possibly also because of the subtle trends in the distribution of cluster masses and ages as a function of galactocentric distance (hence field density). As can be appreciated, the densities involved in the correction for the background (median values for the entire sample) amount to 0.0223 stars per arcsec$^2$ and 324 $L_\sun$ per arcsec$^2$, respectively.

\begin{figure*}
\includegraphics[width=\textwidth]{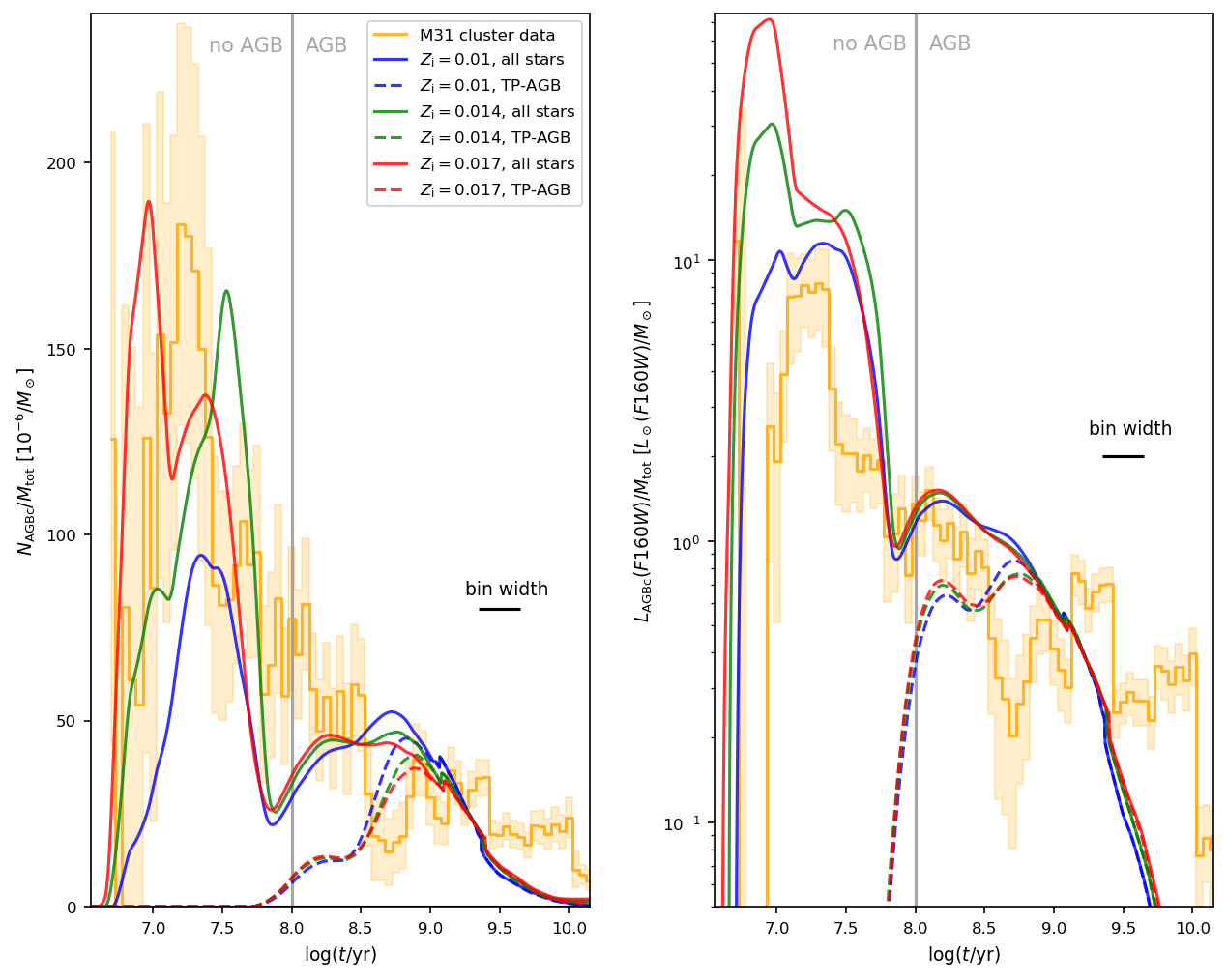}
\caption{The numbers and luminosities of bright red giants per unit mass of stellar populations, at varying age, as derived from our data and compared to stellar population models. The \textbf{left panel} shows the numbers of AGB candidates observed in every age bin, divided by the total mass of all the clusters in the same bin (orange line). Age bins have a fixed width of 0.3-dex in log(age), and are allowed to move over the interval from very young to very old cluster ages at steps of 0.05 dex. The continuous red, green and blue lines show the same quantity,  derived from the PARSEC-COLIBRI model isochrones, for three values of initial metallicity, and again averaged over 0.3-dex wide age intervals. The dashed lines refer to the stars in the TP-AGB section of the isochrones only. The vertical gray line at $\log(t/\mathrm{yr})=8$ separates the broad age ranges where the star counts are expected to be free of AGB stars, and dominated by AGB stars.
The \textbf{right panel} shows, for the same data and sets of models, the total stellar luminosity of bright red giants in the F160W filter, relative to the solar luminosity in the same filter, and per unit mass of stellar populations. In both panels, the filled areas indicate the approximate (and probably underestimated) $1\sigma$ errors of the data.
}
\label{fig_hist_models}
\end{figure*}

Finally, Fig.~\ref{fig_hist_models} shows the $N_\mathrm{AGBc}/M_\mathrm{tot}$ and $L_\mathrm{AGBc}^\mathrm{F160W}/M_\mathrm{tot}$ curves plotted as a function of age. To derive these curves, we adopt the 0.3~dex wide binning in $\log t$, but move the age bins by small steps of 0.05~dex in $\log t$, so as to avoid producing features that depend on the exact location of the bin boundaries.

In addition, we estimate errors for all the quantities involved. Errors in stars counts are taken as the square root of star counts, while errors in cluster masses and F160W luminosities derive directly from the tabulated values. The final errors follow from the propagation of all errors into $N_\mathrm{AGBc}/M_\mathrm{tot}$ and $L_\mathrm{AGBc}^\mathrm{F160W}/M_\mathrm{tot}$, under the assumption that they are independent. Since formal errors in $M_\mathrm{tot}$ are very small (typically less than 10~\% per age bin), errors in $N_\mathrm{AGBc}/M_\mathrm{tot}$ are dominated by the Poisson noise in $N_\mathrm{AGBc}$. Errors in $L_\mathrm{AGBc}^\mathrm{F160W}/M_\mathrm{tot}$, instead, turn out to be unrealistic small when computed this way, since they just reflect the tiny errors in measuring the F160W fluxes of several very bright stars. To include the stochastic fluctuations in the number of AGB candidates into this estimate, we assume the fractional error in $L_\mathrm{AGBc}^\mathrm{F160W}/M_\mathrm{tot}$ to be at least as large as the one in $N_\mathrm{AGBc}/M_\mathrm{tot}$. These approximated (and possibly underestimated) error bars are also plotted in Fig.~\ref{fig_hist_models}. They should be considered as order-of-magnitude estimates of the random errors, only. 

There are other possible errors still not taken into account in the estimates above. For instance, errors related to crowding include a small fraction of AGB candidates ($\sim55$ out of 937) possibly being RGB stars scattered into the AGB region by crowding errors, and the possible blending of a few AGB stars. These errors are not taken into account because they amount to just a few per cent in our present estimates (cf.\ Section \ref{sec_crowd}), which is much less than the errors caused by the Poisson noise in the values of $N_\mathrm{AGBc}$. In addition, there are errors associated with the variable AGB stars being observed at random phases of their pulsation cycles. The latter tend to cancel out for large samples, and should be small for the relatively faint AGB stars located close to the $\mathrm{F160W}$ limit of our sample, which are more likely to be low-amplitude pulsators; on the other hand, they might be significant for the brighter and rarer AGB stars that pulsate as large-amplitude Miras. Therefore, the variability effect could be especially affecting our $L_\mathrm{AGBc}^\mathrm{F160W}/M_\mathrm{tot}$ estimates. However, all these errors are not easy to estimate. We believe that a reliable determination of all the errors in Fig.~\ref{fig_hist_models} can only be reached with a large set of artificial clusters containing variable AGB stars, being inserted into the original PHAT images and having their data recovered exactly as the real clusters.

Finally, one may wonder whether mass segregation between the AGB stars and the other cluster stars that define \Rap, may be affecting our estimates as well. We consider it unlikely since: (1) In theoretical isochrones \citep[e.g, those by][]{marigo17}, AGB stars have initial masses just $\sim10$~\% larger that the turn-off stars, which are the less massive stars that can still contribute significantly to the integrated optical light. Moreover, (2) most of the mass lost by the presently-seen AGB stars was lost in the $\la 2\times10^8$ yr time elapsed since their TRGB (or since the beginning of the core-helium burning phase for intermediate-mass and massive stars) stage. These numbers mean too small a mass difference, and too brief a timespan, for significant mass segregation to take place in the relatively small clusters that make the bulk of our sample.

\subsection{Comparison with models} 
\label{sec_models}

For comparison, in Fig.~\ref{fig_hist_models} we overplot the same quantities $N_\mathrm{AGBc}/M_\mathrm{tot}$ and $L_\mathrm{AGBc}^\mathrm{F160W}/M_\mathrm{tot}$ as derived from the PARSEC-COLIBRI family of isochrones, for initial metallicities between $Z_\mathrm{i}=0.01$ (slightly below solar, $\mh=-0.18$~dex) and $Z_\mathrm{i}=0.017$ (slightly above solar, $\mh=+0.07$~dex), which encompass the median values observed across the M31 disk for galactocentric radii between 4 and 14 kpc \citep[see][]{gregersen15}. These models are available online\footnote{\url{http://stev.oapd.inaf.it/cmd}} and are described in a series of papers including \citet{bressan12}, \citet{chen15}, and \citet{marigo17}. Their TP-AGB sections result from a major effort to provide models calibrated with observations of resolved stellar populations, as described in \citet[][and submitted]{pastorelli19}. The quantities $N_\mathrm{AGBc}/M_\mathrm{tot}$ and $L_\mathrm{AGBc}^\mathrm{F160W}/M_\mathrm{tot}$ are derived by generating, from the isochrones, the photometry of a population with an initial mass of $M_\mathrm{tot}=10^7$~\Msun, located at a true distance modulus of 24.47~mag and with a foreground extinction of $A_V=0.17$~mag. Only stars that satisfy our catalog cut (Eqs.~\ref{eq_magcut} and \ref{eq_colcut}) are then considered in computing $N_\mathrm{AGBc}$ and $L_\mathrm{AGBc}^\mathrm{F160W}$. No attempt is made to include photometric errors in these simulations. 

The $N_\mathrm{AGBc}/M_\mathrm{tot}$ quantity is indicative of the product between the lifetimes of AGB stars, $\tau_\mathrm{AGBc}$, and the rate that stars leave the previous evolutionary stages, $b^\mathrm{AGB}(t)$, as the stellar populations age, that is:
\begin{equation}
    \frac{N_\mathrm{AGBc}}{M_\mathrm{tot}}(t) \propto
        \tau_\mathrm{AGBc}(M_\mathrm{TO}) \times \phi_{M_\mathrm{TO}} \times \left| \frac{d t_\mathrm{eHe}}{d M_\mathrm{i}} \right|^{-1}_{M_\mathrm{i}=M_\mathrm{TO}}  
\end{equation}
\citep[see][]{girardi98}, where $M_\mathrm{TO}$ is the turn-off mass corresponding to the age $t$, $\phi_{M_\mathrm{TO}}$ is the relative number of these stars as predicted by the initial mass function, and $d t_\mathrm{eHe}/d M_\mathrm{i}$ is the rate at which the stellar age at the end of core He-burning phase changes with the initial mass. In general, all these quantities are well-behaved across very wide intervals of $t$, with the only exceptions of $t_\mathrm{eHe}$, which has a major wiggle at 1.6~Gyr \citep[see][]{Girardi_etal13}, and of $\tau_\mathrm{AGBc}$ which is expected to vary a lot as a function of both population age and initial mass, as already said. But given this proportionality, the plot of $N_\mathrm{AGBc}/M_\mathrm{tot}$ should approach zero at the two extremities of the total age interval, since the lifetimes of all cool giants are expected to be zero for the very young clusters containing only blue and yellow supergiants, and becomes very short again at very old ages, as the AGB phase become short-lived and shifts to sub-TRGB luminosities. This is indeed the case, as can be appreciated in Fig.~\ref{fig_hist_models}. We interpret this result as the consequence of performing a good field subtraction in Eq.~\ref{eq_nagb}.

Then, the $N_\mathrm{AGBc}/M_\mathrm{tot}$ curves present a strong peak at ages $7\!<\!\log(t/\mathrm{yr})\!<\!7.6$. The same peak is evident in the stellar population models. They result from the cool evolutionary stages of massive stars, with masses around $\sim12$~\Msun, comprising part of the red core-helium burning phase, and the evolution from central He-exhaustion up to C-ignition in the core. Observationaly, these stars are usually referred to by the generic name of ``red supergiants''. In the models we plot, their numbers generally increase with increasing metallicity. This increase is mainly the result of the Hayashi line systematically moving to cooler effective temperatures as the metallicity increases, hence moving increasing sections of the isochrones to the red of our color cut at $\fJH>0.88$~mag\footnote{Indeed, between the metallicity values we plot in Fig.~\ref{fig_hist_models}, the total lifetimes of red supergiant phases change by factors of just $\sim10$~\% for a given mass.}. The comparison in Fig.~\ref{fig_hist_models} suggests that good agreement between data and models could possibly be obtained at metallicities intermediate between the values $Z_\mathrm{i}=0.01$ and $0.017$. However, reproducing this curve is beyond the scope of the present paper, and it would require a careful consideration of the metallicities sampled in the M31 cluster data. From the point of view of the present work, the main conclusion turning out from the comparison is that this peak in the age distribution of cool giants is well explained by the models, but is \textit{not} related to AGB stars.

The most significant comparison, instead, is for the $8<\log(t/\mathrm{yr})<10.1$ age interval, to the right of the vertical gray line, for which the sample is expected to be dominated by AGB stars. The $N_\mathrm{AGBc}/M_\mathrm{tot}$ curve is observed to decrease almost steadily over this wide age interval, with values going from an initial $\sim50$ stars per million solar masses formed at $10^8$~yr, down to just $\sim5$ at very old ages. But this decrease is marked by significant up-and-down jumps over the entire age interval. The most significant of these features is a temporary decrease in the numbers of stars for ages $8.5<\log(t/\mathrm{yr})<8.8$, where the $N_\mathrm{AGBc}/M_\mathrm{tot}$ curve drops from $\sim40$ down to $\sim20\times10^{-6}\Msun^{-1}$. Later on, as the curve enters in the regime of very old ages ($\log(t/\mathrm{yr})>10$), it drops again from $\sim20$ down to $\sim5\times10^{-6}\Msun^{-1}$. We note that the drop at $8.5<\log(t/\mathrm{yr})<8.8$ coincides with the transition point between CMD ages from \citet{johnson16} and integrated light ages from \citet{beerman15}, so we cannot exclude that it reflects some bias in the ages assigned to the sample.

Comparison with the model values indicate that the predicted numbers of AGB stars are in the observed range, considering the error bars, except for a few discrepancies: 
1) There is no drop predicted for the star counts in the $8.5<\log(t/\mathrm{yr})<8.8$ interval, where model values are at least twice larger than the observed ones. In the models, $N_\mathrm{AGBc}/M_\mathrm{tot}$ actually presents a peak inside this age interval, which is more evident at sub-solar metallicities. This peak occurs mainly because of a steep increase in the lifetimes of TP-AGB stars, just before their maximum values are reached at $\log(t/\mathrm{yr})\simeq9$ \citep[see also][]{GirardiMarigo07,marigo17,pastorelli19}. 
2) At ages $9.2<\log(t/\mathrm{yr})<9.4$, models predict slightly less AGB stars than observed. This age interval corresponds to stars in the TP-AGB phase only, and their numbers present a small dependence on metallicity. It is also the age range in which the ``AGB boosting'' effect appears in the Magellanic Cloud clusters \citep{Girardi_etal13}.
3) Another discrepancy is evident for the oldest clusters, in the bins at ages $\log(t/\mathrm{yr})>9.5$. Theoretical models of single stars predict very few, if any, AGB stars brighther than the TRGB at these ages, while the data seem to suggest their presence at levels between 20 and $5\times10^{-6}\,\Msun^{-1}$. The stars observed could be the progeny of close binaries evolved through the blue straggler phase, but their numbers appear definitely too high considering that the most massive globular clusters in the MW present just a handful of blue stragglers, which become much less likely to be observed after evolving into the AGB. Instead, the problem could be in the uncertainties we already mentioned throughout this paper, which become more serious at the limit of old ages. Indeed, the data for old ages depend on a handful of very massive clusters, with $M_\mathrm{tot}\ga10^5$~\Msun\ (see top-left panel in Fig.~\ref{fig_2dAGBc}), including some of the objects which probably have the most serious problems with crowding. 

\begin{figure*}
\includegraphics[width=\textwidth]{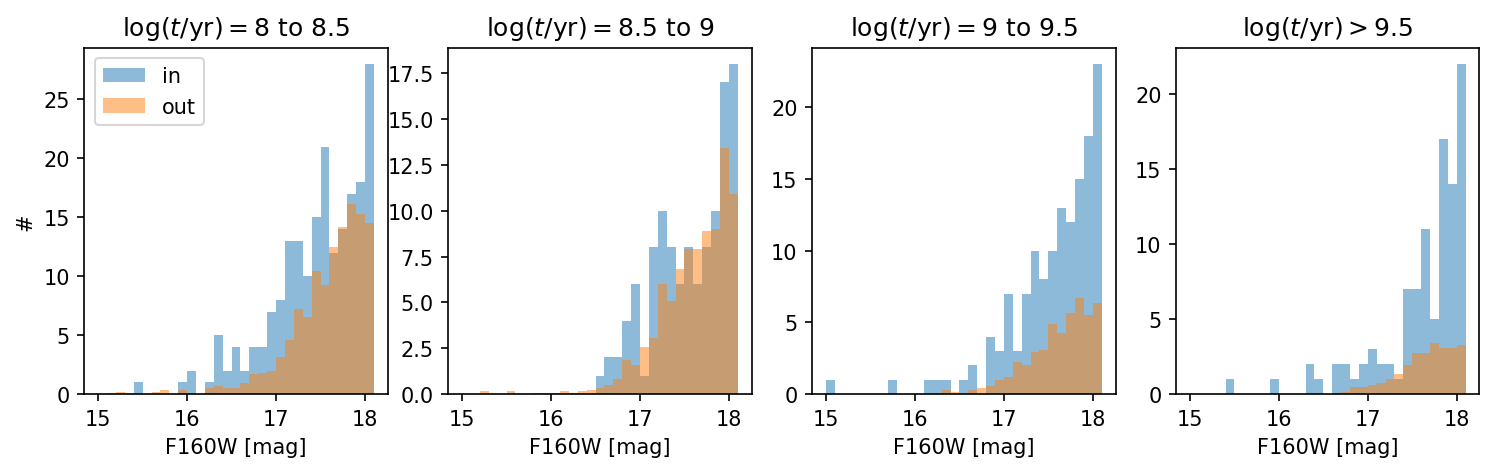}
\caption{Luminosity functions of AGB candidates ('in', for the sample with $\Rsc<\Rap$), separated in several wide age bins. They are compared to the luminosity functions of the stars effectively used as the background sample ('out', for 1/12 of the luminosity function of the sample with $2\Rap<\Rsc<4\Rap$). 
}
\label{fig_f160w_hist}
\end{figure*}

To explore if crowding can really be a serious issue for the oldest clusters, Fig.~\ref{fig_f160w_hist} presents the F160W luminosity functions of candidate AGB stars in the clusters, compared to the luminosity function of the background sample, separated in four wide age intervals. As can be noticed, the properties of the background stars are more or less the same in all age bins, and they are concentrated at $\mathrm{F160W}>17$~mag. For the stars likely to be cluster members (i.e., for the difference between the `in' and `out' histograms in the figure), there is a clear trend of younger age bins having brighter AGB candidates, as expected. They appear as an excess of stars with $16\la\mathrm{F160W}\la17.5$ in the $8<\log(t/\mathrm{yr})<8.5$ age bin, and as an excess at $16.5\la\mathrm{F160W}\la17.5$ in the $8.5<\log(t/\mathrm{yr})<9$ age bin. For the oldest age bins, instead, the distributions are clearly concentrated at the faintest luminosities. Although a peak at the faintest magnitudes ($\mathrm{F160W}\simeq18$~mag) appears for all age bins, this peak is especially remarkable in the oldest age bin, with $\log(t/\mathrm{yr})>9.5$. This serves as a warning that crowding of bright RGB stars could be, for these old clusters, an issue more serious than previously inferred in our previous Sect.~\ref{sec_crowd}. Alternatively, we could be witnessing an unexpected excess of AGB stars in M31 old clusters, or very serious errors in their ages.

Finally, the right-hand panel in Fig.~\ref{fig_hist_models} plots the quantity $L_\mathrm{AGBc}^\mathrm{F160W}/M_\mathrm{tot}$. It is presented here primarily as proxy for the contribution of AGB stars to the near-infrared light of stellar populations \citep[see e.g.][]{melbourne12}. The interpretation of this latter plot is not so obvious. Leaving the discrepancies at very young ages aside, and concentrating on the age interval that corresponds to AGB stars ($\log(t/\mathrm{yr})>8$), we notice the following: First, it is evident that the models reproduce the correct level of $L_\mathrm{AGBc}^\mathrm{F160W}/M_\mathrm{tot}$ in the initial $8.0<\log(t/\mathrm{yr})<8.5$ interval, but for later ages they either underestimate or overestimate the observed luminosities by factors often in excess of 2. These discrepancies partially reflect the discrepancies already noticed in the $N_\mathrm{AGBc}/M_\mathrm{tot}$ plot. The discrepancy seems particularly dramatic as we approach ages $\log(t/\mathrm{yr})>10$ -- where the model values fall to zero, since no stars above the TRGB are predicted at those ages.

To conclude, we invite the readers to appreciate the general level of agreement between the data and the behaviour expected from models, over the very wide age interval under consideration, rather than focusing on the localized discrepancies. Given the all sort of limitations we already mentioned -- such as the high level of field contamination, the great uncertainties possibly affecting cluster ages and aperture radii, the lack of more extensive simulations of effects of crowding, etc. -- this overall agreement was not expected beforehand. It indicates the great potential of these data to offer quantitative constraints to stellar evolution models. 

\section{Closing remarks}
\label{conclu}

In this paper, we present a list of candidate AGB stars in M31 star clusters, derived from PHAT near-infrared photometry added to the AP cluster catalogues built (mainly) from PHAT optical photometry. 

The entire data is made available at the Mikulski Archive for Space
Telescopes\footnote{\url{http://archive.stsci.edu/hlsp/phatagb}} as a High Level Science Product at \dataset[10.17909/t9-mn5h-8e38]{\doi{10.17909/t9-mn5h-8e38}}. It includes a table with all data used in this paper, and ``finding charts'' for all clusters with AGB candidates, as those illustrated in Figs.~\ref{fig_images} to \ref{fig_images_no}. The tabulated data comprise all the quantities already described in \citet{williams14} for the photometry of individual stars, together with the cluster data from the AP catalogue from \citet{johnson15}, and their age and mass estimates computed as in \citet{Fouesneau_etal14} and \citet{beerman15}. To all these data, we add just a few quantities derived in this paper, like the distance of stars from the cluster centre (\Rsc; Sect.~\ref{sec_match}), a flag to point stars potentially misclassified by crowding, and the RGB density (Sect.~\ref{sec_crowd}). We also add the spectral classification (Sects.~\ref{sec_CM} and \ref{sec_spectra}) and a variability flag (Sect.~\ref{sec_variables}), for the stars for which this information is available.

Out of an initial list of 937 candidates, nearly half (467) are likely field stars appearing by chance within the cluster aperture radius (Sect.~\ref{data}). Moreover, out of the 791 candidates with valid cluster ages, 295 belong to clusters that are too young to host an AGB population. These stars are probably not AGB stars but younger ``red supergiants'' of different kinds (Sect.~\ref{results}). Taking these fractions into consideration, our catalogue contains $\sim294$ stars which are very likely AGB members of star clusters in M31. This number exceeds by a factor of $\sim2$ any other list of AGB stars in clusters, including those for the Magellanic Clouds which have driven so much progress in our understanding of this evolutionary phase. 

Therefore, the present sample of candidate AGB stars might be a treasure trove for future studies of stellar evolution. For instance, they might help to set clearer constraints on the mass and age intervals that allow for the formation of C and S stars, and their luminosities, and allow a more direct assessment of their evolutionary lifetimes. These constraints are badly needed at the high metallicities that characterize M31 and the MW galaxy. Similar studies in the MW clusters do not appear that promising, given the small samples of stellar clusters available in optical surveys (from where ages are better determined), and their small masses. Although Gaia, the Rubin Observatory Legacy Survey of Space and Time, and future wide-area high-resolution NIR surveys (e.g., with the Nancy Grace Roman Space Telescope) might improve this situation, having a sample of clusters observed from outside, and with a good understanding of their contaminating foreground/background like in M31, will certainly help. 

Many of the population studies enabled by the present catalogue will, however, require follow-up observations of some kind, and more evidently: (1) spectroscopy to determine the spectral types of the TP-AGB candidates (broadly speaking, into M, C, S, dust-obscured, and CN-weak subtypes, see e.g., \citealt{hamren15} and the examples in Sect.~\ref{sec_spectra}), and hopefully their radial velocities compared to the clusters for membership probabilities; or alternatively (2) narrow/medium-band photometry for a more efficient C/M classification \citep[see][and Sect.~\ref{sec_CM}]{boyer13,boyer19}; (3) time series photometry to identify long-period variables (as in Sect.~\ref{sec_variables}) and to determine their pulsation periods and modes, either with adaptive optics or with future space-based NIR facilities like the James Webb and Roman space telescopes; (4) higher resolution, deeper optical photometry for improving cluster age determinations. The present catalogues are released especially to motivate such follow-up efforts. 

However, it is also clear that the M31 sample, as a whole, will be much more affected by the problems of field contamination and age uncertainties than the AGB star samples already studied in the Magellanic Clouds clusters. They will therefore require careful analysis of membership probabilities and simulations of measurement errors in the stellar photometry, and in the cluster ages and masses. These aspects will be more thoroughly discussed in a forthcoming paper, where the M31 sample will be compared with a similar sample from M33 (HST-GO-14610, PI: Dalcanton).

\acknowledgments
This research is based on observations made with the NASA/ESA Hubble Space Telescope obtained from the Space Telescope Science Institute, which is operated by the Association of Universities for Research in Astronomy, Inc., under NASA contract NAS 5–26555. These observations are associated with HST programs GO-12055 and GO-14072. We acknowledge the support from the ERC Consolidator Grant funding scheme ({\em project STARKEY}, G.A. n. 615604). We thank the anonymous referee for the detailed report and useful suggestions.

%

\vspace{5mm}
\facilities{HST (ACS), HST (WFC3), Keck (DEIMOS)}


\software{Perl \citep{perl}, STILTS \citep{stilts}, astropy \citep{astropy}, DOLPHOT \citep{dolphot}}





\bibliography{AGBclust}{}

\begin{thebibliography}{}
\expandafter\ifx\csname natexlab\endcsname\relax\def\natexlab#1{#1}\fi
\providecommand{\url}[1]{\href{#1}{#1}}
\providecommand{\dodoi}[1]{doi:~\href{http://doi.org/#1}{\nolinkurl{#1}}}
\providecommand{\doeprint}[1]{\href{http://ascl.net/#1}{\nolinkurl{http://ascl.net/#1}}}
\providecommand{\doarXiv}[1]{\href{https://arxiv.org/abs/#1}{\nolinkurl{https://arxiv.org/abs/#1}}}

\bibitem[{{Astropy Collaboration} {et~al.}(2013){Astropy Collaboration},
  {Robitaille}, {Tollerud}, {Greenfield}, {Droettboom}, {Bray}, {Aldcroft},
  {Davis}, {Ginsburg}, {Price-Whelan}, {Kerzendorf}, {Conley}, {Crighton},
  {Barbary}, {Muna}, {Ferguson}, {Grollier}, {Parikh}, {Nair}, {Unther},
  {Deil}, {Woillez}, {Conseil}, {Kramer}, {Turner}, {Singer}, {Fox}, {Weaver},
  {Zabalza}, {Edwards}, {Azalee Bostroem}, {Burke}, {Casey}, {Crawford},
  {Dencheva}, {Ely}, {Jenness}, {Labrie}, {Lim}, {Pierfederici}, {Pontzen},
  {Ptak}, {Refsdal}, {Servillat}, \& {Streicher}}]{astropy}
{Astropy Collaboration}, {Robitaille}, T.~P., {Tollerud}, E.~J., {et~al.} 2013,
  \aap, 558, A33, \dodoi{10.1051/0004-6361/201322068}

\bibitem[{{Beerman}(2015)}]{beerman15}
{Beerman}, L.~C. 2015, PhD thesis, University of Washington

\bibitem[{{Bernard} {et~al.}(2015){Bernard}, {Ferguson}, {Chapman}, {Ibata},
  {Irwin}, {Lewis}, \& {McConnachie}}]{Bernard_etal14}
{Bernard}, E.~J., {Ferguson}, A. M.~N., {Chapman}, S.~C., {et~al.} 2015,
  \mnras, 453, L113, \dodoi{10.1093/mnrasl/slv116}

\bibitem[{{Bernard} {et~al.}(2012){Bernard}, {Ferguson}, {Barker}, {Hidalgo},
  {Ibata}, {Irwin}, {Lewis}, {McConnachie}, {Monelli}, \&
  {Chapman}}]{Bernard_etal12}
{Bernard}, E.~J., {Ferguson}, A.~M.~N., {Barker}, M.~K., {et~al.} 2012, \mnras,
  420, 2625, \dodoi{10.1111/j.1365-2966.2011.20234.x}

\bibitem[{{Boyer} {et~al.}(2008){Boyer}, {McDonald}, {Loon}, {Woodward},
  {Gehrz}, {Evans}, \& {Dupree}}]{Boyer_etal08}
{Boyer}, M.~L., {McDonald}, I., {Loon}, J.~T., {et~al.} 2008, \aj, 135, 1395,
  \dodoi{10.1088/0004-6256/135/4/1395}

\bibitem[{{Boyer} {et~al.}(2010){Boyer}, {van Loon}, {McDonald}, {Gordon},
  {Babler}, {Block}, {Bracker}, {Engelbracht}, {Hora}, {Indebetouw}, {Meade},
  {Meixner}, {Misselt}, {Sewilo}, {Shiao}, \& {Whitney}}]{boyer10}
{Boyer}, M.~L., {van Loon}, J.~T., {McDonald}, I., {et~al.} 2010, \apjl, 711,
  L99, \dodoi{10.1088/2041-8205/711/2/L99}

\bibitem[{{Boyer} {et~al.}(2011){Boyer}, {Srinivasan}, {van Loon}, {McDonald},
  {Meixner}, {Zaritsky}, {Gordon}, {Kemper}, {Babler}, {Block}, {Bracker},
  {Engelbracht}, {Hora}, {Indebetouw}, {Meade}, {Misselt}, {Robitaille},
  {Sewi{\l}o}, {Shiao}, \& {Whitney}}]{boyer11}
{Boyer}, M.~L., {Srinivasan}, S., {van Loon}, J.~T., {et~al.} 2011, \aj, 142,
  103, \dodoi{10.1088/0004-6256/142/4/103}

\bibitem[{{Boyer} {et~al.}(2013){Boyer}, {Girardi}, {Marigo}, {Williams},
  {Aringer}, {Nowotny}, {Rosenfield}, {Dorman}, {Guhathakurta}, {Dalcanton},
  {Melbourne}, {Olsen}, \& {Weisz}}]{boyer13}
{Boyer}, M.~L., {Girardi}, L., {Marigo}, P., {et~al.} 2013, \apj, 774, 83,
  \dodoi{10.1088/0004-637X/774/1/83}

\bibitem[{{Boyer} {et~al.}(2019){Boyer}, {Williams}, {Aringer}, {Chen},
  {Dalcanton}, {Girardi}, {Guhathakurta}, {Marigo}, {Olsen}, {Rosenfield}, \&
  {Weisz}}]{boyer19}
{Boyer}, M.~L., {Williams}, B.~F., {Aringer}, B., {et~al.} 2019, \apj, 879,
  109, \dodoi{10.3847/1538-4357/ab24e2}

\bibitem[{{Bressan} {et~al.}(2012){Bressan}, {Marigo}, {Girardi}, {Salasnich},
  {Dal Cero}, {Rubele}, \& {Nanni}}]{bressan12}
{Bressan}, A., {Marigo}, P., {Girardi}, L., {et~al.} 2012, \mnras, 427, 127,
  \dodoi{10.1111/j.1365-2966.2012.21948.x}

\bibitem[{{Brown} {et~al.}(2006){Brown}, {Smith}, {Ferguson}, {Rich},
  {Guhathakurta}, {Renzini}, {Sweigart}, \& {Kimble}}]{brown06}
{Brown}, T.~M., {Smith}, E., {Ferguson}, H.~C., {et~al.} 2006, \apj, 652, 323,
  \dodoi{10.1086/508015}

\bibitem[{{Caldwell} {et~al.}(2009){Caldwell}, {Harding}, {Morrison}, {Rose},
  {Schiavon}, \& {Kriessler}}]{caldwell09}
{Caldwell}, N., {Harding}, P., {Morrison}, H., {et~al.} 2009, \aj, 137, 94,
  \dodoi{10.1088/0004-6256/137/1/94}

\bibitem[{{Catchpole} \& {Feast}(1973)}]{catchpole73}
{Catchpole}, R.~M., \& {Feast}, M.~W. 1973, \mnras, 164, 11P,
  \dodoi{10.1093/mnras/164.1.11P}

\bibitem[{{Chen} {et~al.}(2015){Chen}, {Bressan}, {Girardi}, {Marigo}, {Kong},
  \& {Lanza}}]{chen15}
{Chen}, Y., {Bressan}, A., {Girardi}, L., {et~al.} 2015, \mnras, 452, 1068,
  \dodoi{10.1093/mnras/stv1281}

\bibitem[{{Cioni} {et~al.}(2006{\natexlab{a}}){Cioni}, {Girardi}, {Marigo}, \&
  {Habing}}]{cioni06lmc}
{Cioni}, M.-R.~L., {Girardi}, L., {Marigo}, P., \& {Habing}, H.~J.
  2006{\natexlab{a}}, \aap, 448, 77, \dodoi{10.1051/0004-6361:20053933}

\bibitem[{{Cioni} {et~al.}(2006{\natexlab{b}}){Cioni}, {Girardi}, {Marigo}, \&
  {Habing}}]{cioni06smc}
---. 2006{\natexlab{b}}, \aap, 452, 195, \dodoi{10.1051/0004-6361:20054699}

\bibitem[{{Cioni} {et~al.}(2011){Cioni}, {Clementini}, {Girardi}, {Guandalini},
  {Gullieuszik}, {Miszalski}, {Moretti}, {Ripepi}, {Rubele}, {Bagheri},
  {Bekki}, {Cross}, {de Blok}, {de Grijs}, {Emerson}, {Evans}, {Gibson},
  {Gonzales-Solares}, {Groenewegen}, {Irwin}, {Ivanov}, {Lewis}, {Marconi},
  {Marquette}, {Mastropietro}, {Moore}, {Napiwotzki}, {Naylor}, {Oliveira},
  {Read}, {Sutorius}, {van Loon}, {Wilkinson}, \& {Wood}}]{Cioni_etal11}
{Cioni}, M.-R.~L., {Clementini}, G., {Girardi}, L., {et~al.} 2011, \aap, 527,
  A116, \dodoi{10.1051/0004-6361/201016137}

\bibitem[{{Dalcanton} {et~al.}(2012){Dalcanton}, {Williams}, {Lang}, {Lauer},
  {Kalirai}, {Seth}, {Dolphin}, {Rosenfield}, {Weisz}, {Bell}, {Bianchi},
  {Boyer}, {Caldwell}, {Dong}, {Dorman}, {Gilbert}, {Girardi}, {Gogarten},
  {Gordon}, {Guhathakurta}, {Hodge}, {Holtzman}, {Johnson}, {Larsen}, {Lewis},
  {Melbourne}, {Olsen}, {Rix}, {Rosema}, {Saha}, {Sarajedini}, {Skillman}, \&
  {Stanek}}]{Dalcanton_etal12}
{Dalcanton}, J.~J., {Williams}, B.~F., {Lang}, D., {et~al.} 2012, \apjs, 200,
  18, \dodoi{10.1088/0067-0049/200/2/18}

\bibitem[{{Davidge}(2012)}]{davidge12}
{Davidge}, T.~J. 2012, \apjl, 749, L7, \dodoi{10.1088/2041-8205/749/1/L7}

\bibitem[{{Davis} {et~al.}(2019){Davis}, {Bond}, {Ciardullo}, \&
  {Jacoby}}]{davis19}
{Davis}, B.~D., {Bond}, H.~E., {Ciardullo}, R., \& {Jacoby}, G.~H. 2019, \apj,
  884, 115, \dodoi{10.3847/1538-4357/ab44d4}

\bibitem[{{Dolphin}(2016)}]{dolphot}
{Dolphin}, A. 2016, {DOLPHOT: Stellar photometry}.
\newblock \doeprint{1608.013}

\bibitem[{{Dorman} {et~al.}(2012){Dorman}, {Guhathakurta}, {Fardal}, {Lang},
  {Geha}, {Howley}, {Kalirai}, {Bullock}, {Cuilland re}, {Dalcanton},
  {Gilbert}, {Seth}, {Tollerud}, {Williams}, \& {Yniguez}}]{dorman12}
{Dorman}, C.~E., {Guhathakurta}, P., {Fardal}, M.~A., {et~al.} 2012, \apj, 752,
  147, \dodoi{10.1088/0004-637X/752/2/147}

\bibitem[{{Dorman} {et~al.}(2013){Dorman}, {Widrow}, {Guhathakurta}, {Seth},
  {Foreman-Mackey}, {Bell}, {Dalcanton}, {Gilbert}, {Skillman}, \&
  {Williams}}]{dorman13}
{Dorman}, C.~E., {Widrow}, L.~M., {Guhathakurta}, P., {et~al.} 2013, \apj, 779,
  103, \dodoi{10.1088/0004-637X/779/2/103}

\bibitem[{{Dorman} {et~al.}(2015){Dorman}, {Guhathakurta}, {Seth}, {Weisz},
  {Bell}, {Dalcanton}, {Gilbert}, {Hamren}, {Lewis}, {Skillman}, {Toloba}, \&
  {Williams}}]{dorman15}
{Dorman}, C.~E., {Guhathakurta}, P., {Seth}, A.~C., {et~al.} 2015, \apj, 803,
  24, \dodoi{10.1088/0004-637X/803/1/24}

\bibitem[{{Ferguson} {et~al.}(2005){Ferguson}, {Johnson}, {Faria}, {Irwin},
  {Ibata}, {Johnston}, {Lewis}, \& {Tanvir}}]{ferguson05}
{Ferguson}, A. M.~N., {Johnson}, R.~A., {Faria}, D.~C., {et~al.} 2005, \apjl,
  622, L109, \dodoi{10.1086/429371}

\bibitem[{{Ferraro} {et~al.}(1999){Ferraro}, {Messineo}, {Fusi Pecci}, {de
  Palo}, {Straniero}, {Chieffi}, \& {Limongi}}]{ferraro99}
{Ferraro}, F.~R., {Messineo}, M., {Fusi Pecci}, F., {et~al.} 1999, \aj, 118,
  1738, \dodoi{10.1086/301029}

\bibitem[{{Fouesneau} {et~al.}(2014){Fouesneau}, {Johnson}, {Weisz},
  {Dalcanton}, {Bell}, {Bianchi}, {Caldwell}, {Gouliermis}, {Guhathakurta},
  {Kalirai}, {Larsen}, {Rix}, {Seth}, {Skillman}, \&
  {Williams}}]{Fouesneau_etal14}
{Fouesneau}, M., {Johnson}, L.~C., {Weisz}, D.~R., {et~al.} 2014, \apj, 786,
  117, \dodoi{10.1088/0004-637X/786/2/117}

\bibitem[{{Frogel} {et~al.}(1990){Frogel}, {Mould}, \&
  {Blanco}}]{Frogel_etal90}
{Frogel}, J.~A., {Mould}, J., \& {Blanco}, V.~M. 1990, \apj, 352, 96,
  \dodoi{10.1086/168518}

\bibitem[{{Galleti} {et~al.}(2004){Galleti}, {Federici}, {Bellazzini}, {Fusi
  Pecci}, \& {Macrina}}]{bolognacat}
{Galleti}, S., {Federici}, L., {Bellazzini}, M., {Fusi Pecci}, F., \&
  {Macrina}, S. 2004, \aap, 416, 917, \dodoi{10.1051/0004-6361:20035632}

\bibitem[{{Gaustad} \& {Conti}(1971)}]{gaustad71}
{Gaustad}, J.~E., \& {Conti}, P.~S. 1971, \pasp, 83, 351,
  \dodoi{10.1086/129135}

\bibitem[{{Gilbert} {et~al.}(2006){Gilbert}, {Guhathakurta}, {Kalirai}, {Rich},
  {Majewski}, {Ostheimer}, {Reitzel}, {Cenarro}, {Cooper}, {Luine}, \&
  {Patterson}}]{gilbert06}
{Gilbert}, K.~M., {Guhathakurta}, P., {Kalirai}, J.~S., {et~al.} 2006, \apj,
  652, 1188, \dodoi{10.1086/508643}

\bibitem[{{Girardi} \& {Bertelli}(1998)}]{girardi98}
{Girardi}, L., \& {Bertelli}, G. 1998, \mnras, 300, 533,
  \dodoi{10.1046/j.1365-8711.1998.01934.x}

\bibitem[{{Girardi} \& {Marigo}(2007)}]{GirardiMarigo07}
{Girardi}, L., \& {Marigo}, P. 2007, \aap, 462, 237,
  \dodoi{10.1051/0004-6361:20065249}

\bibitem[{{Girardi} {et~al.}(2013){Girardi}, {Marigo}, {Bressan}, \&
  {Rosenfield}}]{Girardi_etal13}
{Girardi}, L., {Marigo}, P., {Bressan}, A., \& {Rosenfield}, P. 2013, \apj,
  777, 142, \dodoi{10.1088/0004-637X/777/2/142}

\bibitem[{{Gregersen} {et~al.}(2015){Gregersen}, {Seth}, {Williams}, {Lang},
  {Dalcanton}, {Girardi}, {Skillman}, {Bell}, {Dolphin}, {Fouesneau},
  {Guhathakurta}, {Hamren}, {Johnson}, {Kalirai}, {Lewis}, {Monachesi}, \&
  {Olsen}}]{gregersen15}
{Gregersen}, D., {Seth}, A.~C., {Williams}, B.~F., {et~al.} 2015, \aj, 150,
  189, \dodoi{10.1088/0004-6256/150/6/189}

\bibitem[{{Guhathakurta} {et~al.}(2005){Guhathakurta}, {Ostheimer}, {Gilbert},
  {Rich}, {Majewski}, {Kalirai}, {Reitzel}, \& {Patterson}}]{raja05}
{Guhathakurta}, P., {Ostheimer}, J.~C., {Gilbert}, K.~M., {et~al.} 2005, arXiv
  e-prints, astro.
\newblock \doarXiv{astro-ph/0502366}

\bibitem[{{Guhathakurta} {et~al.}(2006){Guhathakurta}, {Rich}, {Reitzel},
  {Cooper}, {Gilbert}, {Majewski}, {Ostheimer}, {Geha}, {Johnston}, \&
  {Patterson}}]{raja06}
{Guhathakurta}, P., {Rich}, R.~M., {Reitzel}, D.~B., {et~al.} 2006, \aj, 131,
  2497, \dodoi{10.1086/499562}

\bibitem[{{Hamren} {et~al.}(2015){Hamren}, {Rockosi}, {Guhathakurta}, {Boyer},
  {Smith}, {Dalcanton}, {Gregersen}, {Seth}, {Lewis}, {Williams}, {Toloba},
  {Girardi}, {Dorman}, {Gilbert}, \& {Weisz}}]{hamren15}
{Hamren}, K.~M., {Rockosi}, C.~M., {Guhathakurta}, P., {et~al.} 2015, \apj,
  810, 60, \dodoi{10.1088/0004-637X/810/1/60}

\bibitem[{{Hartwick} \& {Hesser}(1973)}]{hartwick73}
{Hartwick}, F.~D.~A., \& {Hesser}, J.~E. 1973, \apj, 183, 883,
  \dodoi{10.1086/152275}

\bibitem[{{Herwig}(2005)}]{herwig05}
{Herwig}, F. 2005, \araa, 43, 435,
  \dodoi{10.1146/annurev.astro.43.072103.150600}

\bibitem[{{H{\"o}fner} \& {Olofsson}(2018)}]{hoefner18}
{H{\"o}fner}, S., \& {Olofsson}, H. 2018, \aapr, 26, 1,
  \dodoi{10.1007/s00159-017-0106-5}

\bibitem[{{Javadi} {et~al.}(2011){Javadi}, {van Loon}, \&
  {Mirtorabi}}]{Javadi_etal11}
{Javadi}, A., {van Loon}, J.~T., \& {Mirtorabi}, M.~T. 2011, \mnras, 414, 3394,
  \dodoi{10.1111/j.1365-2966.2011.18638.x}

\bibitem[{{Johnson} {et~al.}(2012){Johnson}, {Seth}, {Dalcanton}, {Caldwell},
  {Fouesneau}, {Gouliermis}, {Hodge}, {Larsen}, {Olsen}, {San Roman},
  {Sarajedini}, {Weisz}, {Williams}, {Beerman}, {Bianchi}, {Dolphin},
  {Girardi}, {Guhathakurta}, {Kalirai}, {Lang}, {Monachesi}, {Nanda}, {Rix}, \&
  {Skillman}}]{Johnson_etal12}
{Johnson}, L.~C., {Seth}, A.~C., {Dalcanton}, J.~J., {et~al.} 2012, \apj, 752,
  95, \dodoi{10.1088/0004-637X/752/2/95}

\bibitem[{{Johnson} {et~al.}(2015){Johnson}, {Seth}, {Dalcanton}, {Wallace},
  {Simpson}, {Lintott}, {Kapadia}, {Skillman}, {Caldwell}, {Fouesneau},
  {Weisz}, {Williams}, {Beerman}, {Gouliermis}, \& {Sarajedini}}]{johnson15}
---. 2015, \apj, 802, 127, \dodoi{10.1088/0004-637X/802/2/127}

\bibitem[{{Johnson} {et~al.}(2016){Johnson}, {Seth}, {Dalcanton}, {Beerman},
  {Fouesneau}, {Lewis}, {Weisz}, {Williams}, {Bell}, {Dolphin}, {Larsen},
  {Sandstrom}, \& {Skillman}}]{johnson16}
---. 2016, \apj, 827, 33, \dodoi{10.3847/0004-637X/827/1/33}

\bibitem[{{Kalinowski} {et~al.}(1974){Kalinowski}, {Burkhead}, \&
  {Honeycutt}}]{kalinowski74}
{Kalinowski}, J.~K., {Burkhead}, M.~S., \& {Honeycutt}, R.~K. 1974, \apjl, 193,
  L77, \dodoi{10.1086/181636}

\bibitem[{{Kamath} {et~al.}(2012){Kamath}, {Karakas}, \&
  {Wood}}]{Kamath_etal12}
{Kamath}, D., {Karakas}, A.~I., \& {Wood}, P.~R. 2012, \apj, 746, 20,
  \dodoi{10.1088/0004-637X/746/1/20}

\bibitem[{{Kamath} {et~al.}(2010){Kamath}, {Wood}, {Soszy{\'n}ski}, \&
  {Lebzelter}}]{Kamath_etal10}
{Kamath}, D., {Wood}, P.~R., {Soszy{\'n}ski}, I., \& {Lebzelter}, T. 2010,
  \mnras, 408, 522, \dodoi{10.1111/j.1365-2966.2010.17137.x}

\bibitem[{{Lattanzio} \& {Wood}(2004)}]{lattanzio04}
{Lattanzio}, J.~C., \& {Wood}, P.~R. 2004, {Evolution, Nucleosynthesis, and
  Pulsation of AGB Stars}, 23--104, \dodoi{10.1007/978-1-4757-3876-6_2}

\bibitem[{{Lebzelter} {et~al.}(2008){Lebzelter}, {Lederer}, {Cristallo},
  {Hinkle}, {Straniero}, \& {Aringer}}]{Lebzelter_etal08}
{Lebzelter}, T., {Lederer}, M.~T., {Cristallo}, S., {et~al.} 2008, \aap, 486,
  511, \dodoi{10.1051/0004-6361:200809363}

\bibitem[{{Lebzelter} {et~al.}(2014){Lebzelter}, {Nowotny}, {Hinkle},
  {H{\"o}fner}, \& {Aringer}}]{Lebzelter_etal14}
{Lebzelter}, T., {Nowotny}, W., {Hinkle}, K.~H., {H{\"o}fner}, S., \&
  {Aringer}, B. 2014, \aap, 567, A143, \dodoi{10.1051/0004-6361/201424078}

\bibitem[{{Lebzelter} \& {Wood}(2005)}]{LebzelterWood05}
{Lebzelter}, T., \& {Wood}, P.~R. 2005, \aap, 441, 1117,
  \dodoi{10.1051/0004-6361:20053464}

\bibitem[{{Lebzelter} \& {Wood}(2011)}]{LebzelterWood11}
---. 2011, \aap, 529, A137, \dodoi{10.1051/0004-6361/201016319}

\bibitem[{{Lewis} {et~al.}(2015){Lewis}, {Dolphin}, {Dalcanton}, {Weisz},
  {Williams}, {Bell}, {Seth}, {Simones}, {Skillman}, {Choi}, {Fouesneau},
  {Guhathakurta}, {Johnson}, {Kalirai}, {Leroy}, {Monachesi}, {Rix}, \&
  {Schruba}}]{lewis15}
{Lewis}, A.~R., {Dolphin}, A.~E., {Dalcanton}, J.~J., {et~al.} 2015, \apj, 805,
  183, \dodoi{10.1088/0004-637X/805/2/183}

\bibitem[{{Maraston}(2005)}]{Maraston05}
{Maraston}, C. 2005, \mnras, 362, 799, \dodoi{10.1111/j.1365-2966.2005.09270.x}

\bibitem[{{Marigo} \& {Girardi}(2007)}]{marigo07}
{Marigo}, P., \& {Girardi}, L. 2007, \aap, 469, 239,
  \dodoi{10.1051/0004-6361:20066772}

\bibitem[{{Marigo} {et~al.}(1996){Marigo}, {Girardi}, \&
  {Chiosi}}]{Marigo_etal96}
{Marigo}, P., {Girardi}, L., \& {Chiosi}, C. 1996, \aap, 316, L1

\bibitem[{{Marigo} {et~al.}(2017){Marigo}, {Girardi}, {Bressan}, {Rosenfield},
  {Aringer}, {Chen}, {Dussin}, {Nanni}, {Pastorelli}, {Rodrigues}, {Trabucchi},
  {Bladh}, {Dalcanton}, {Groenewegen}, {Montalb{\'a}n}, \& {Wood}}]{marigo17}
{Marigo}, P., {Girardi}, L., {Bressan}, A., {et~al.} 2017, \apj, 835, 77,
  \dodoi{10.3847/1538-4357/835/1/77}

\bibitem[{{McDonald} {et~al.}(2011{\natexlab{a}}){McDonald}, {Boyer}, {van
  Loon}, \& {Zijlstra}}]{mcdonald11}
{McDonald}, I., {Boyer}, M.~L., {van Loon}, J.~T., \& {Zijlstra}, A.~A.
  2011{\natexlab{a}}, \apj, 730, 71, \dodoi{10.1088/0004-637X/730/2/71}

\bibitem[{{McDonald} {et~al.}(2009){McDonald}, {van Loon}, {Decin}, {Boyer},
  {Dupree}, {Evans}, {Gehrz}, \& {Woodward}}]{McDonald_etal09}
{McDonald}, I., {van Loon}, J.~T., {Decin}, L., {et~al.} 2009, \mnras, 394,
  831, \dodoi{10.1111/j.1365-2966.2008.14370.x}

\bibitem[{{McDonald} {et~al.}(2011{\natexlab{b}}){McDonald}, {van Loon},
  {Sloan}, {Dupree}, {Zijlstra}, {Boyer}, {Gehrz}, {Evans}, {Woodward}, \&
  {Johnson}}]{McDonald_etal11}
{McDonald}, I., {van Loon}, J.~T., {Sloan}, G.~C., {et~al.} 2011{\natexlab{b}},
  \mnras, 417, 20, \dodoi{10.1111/j.1365-2966.2011.18963.x}

\bibitem[{{Melbourne} {et~al.}(2012){Melbourne}, {Williams}, {Dalcanton},
  {Rosenfield}, {Girardi}, {Marigo}, {Weisz}, {Dolphin}, {Boyer}, {Olsen},
  {Skillman}, \& {Seth}}]{melbourne12}
{Melbourne}, J., {Williams}, B.~F., {Dalcanton}, J.~J., {et~al.} 2012, \apj,
  748, 47, \dodoi{10.1088/0004-637X/748/1/47}

\bibitem[{{Momany} {et~al.}(2012){Momany}, {Saviane}, {Smette}, {Bayo},
  {Girardi}, {Marconi}, {Milone}, \& {Bressan}}]{momany12}
{Momany}, Y., {Saviane}, I., {Smette}, A., {et~al.} 2012, \aap, 537, A2,
  \dodoi{10.1051/0004-6361/201117223}

\bibitem[{{No{\"e}l} {et~al.}(2013){No{\"e}l}, {Greggio}, {Renzini}, {Carollo},
  \& {Maraston}}]{Noel_etal13}
{No{\"e}l}, N.~E.~D., {Greggio}, L., {Renzini}, A., {Carollo}, C.~M., \&
  {Maraston}, C. 2013, \apj, 772, 58, \dodoi{10.1088/0004-637X/772/1/58}

\bibitem[{{Pastorelli} {et~al.}(2019){Pastorelli}, {Marigo}, {Girardi}, {Chen},
  {Rubele}, {Trabucchi}, {Aringer}, {Bladh}, {Bressan}, {Montalb{\'a}n},
  {Boyer}, {Dalcanton}, {Eriksson}, {Groenewegen}, {H{\"o}fner}, {Lebzelter},
  {Nanni}, {Rosenfield}, {Wood}, \& {Cioni}}]{pastorelli19}
{Pastorelli}, G., {Marigo}, P., {Girardi}, L., {et~al.} 2019, \mnras, 485,
  5666, \dodoi{10.1093/mnras/stz725}

\bibitem[{{Peacock} {et~al.}(2010){Peacock}, {Maccarone}, {Knigge}, {Kundu},
  {Waters}, {Zepf}, \& {Zurek}}]{peacock10}
{Peacock}, M.~B., {Maccarone}, T.~J., {Knigge}, C., {et~al.} 2010, \mnras, 402,
  803, \dodoi{10.1111/j.1365-2966.2009.15952.x}

\bibitem[{{Pessev} {et~al.}(2008){Pessev}, {Goudfrooij}, {Puzia}, \&
  {Chandar}}]{Pessev_etal08}
{Pessev}, P.~M., {Goudfrooij}, P., {Puzia}, T.~H., \& {Chandar}, R. 2008,
  \mnras, 385, 1535, \dodoi{10.1111/j.1365-2966.2008.12935.x}

\bibitem[{{Quirk} {et~al.}(2019){Quirk}, {Guhathakurta}, {Chemin}, {Dorman},
  {Gilbert}, {Seth}, {Williams}, \& {Dalcanton}}]{quirk19}
{Quirk}, A., {Guhathakurta}, P., {Chemin}, L., {et~al.} 2019, \apj, 871, 11,
  \dodoi{10.3847/1538-4357/aaf1ba}

\bibitem[{{San Roman} {et~al.}(2010){San Roman}, {Sarajedini}, \&
  {Aparicio}}]{sanroman10}
{San Roman}, I., {Sarajedini}, A., \& {Aparicio}, A. 2010, \apj, 720, 1674,
  \dodoi{10.1088/0004-637X/720/2/1674}

\bibitem[{{Santos} \& {Frogel}(1997)}]{SantosFrogel97}
{Santos}, Jr., J.~F.~C., \& {Frogel}, J.~A. 1997, \apj, 479, 764

\bibitem[{{Sarajedini} \& {Mancone}(2007)}]{sarajedini07}
{Sarajedini}, A., \& {Mancone}, C.~L. 2007, \aj, 134, 447,
  \dodoi{10.1086/518835}

\bibitem[{{Schweizer} {et~al.}(1996){Schweizer}, {Miller}, {Whitmore}, \&
  {Fall}}]{schweizer96}
{Schweizer}, F., {Miller}, B.~W., {Whitmore}, B.~C., \& {Fall}, S.~M. 1996,
  \aj, 112, 1839, \dodoi{10.1086/118146}

\bibitem[{{Senchyna} {et~al.}(2015){Senchyna}, {Johnson}, {Dalcanton},
  {Beerman}, {Fouesneau}, {Dolphin}, {Williams}, {Rosenfield}, \&
  {Larsen}}]{senchyna15}
{Senchyna}, P., {Johnson}, L.~C., {Dalcanton}, J.~J., {et~al.} 2015, \apj, 813,
  31, \dodoi{10.1088/0004-637X/813/1/31}

\bibitem[{{Soraisam} {et~al.}(2020){Soraisam}, {Bildsten}, {Drout}, {Prince},
  {Kupfer}, {Masci}, {Laher}, \& {Kulkarni}}]{soraisam20}
{Soraisam}, M.~D., {Bildsten}, L., {Drout}, M.~R., {et~al.} 2020, \apj, 893,
  11, \dodoi{10.3847/1538-4357/ab7b7b}

\bibitem[{{Taylor}(2006)}]{stilts}
{Taylor}, M.~B. 2006, in Astronomical Society of the Pacific Conference Series,
  Vol. 351, Astronomical Data Analysis Software and Systems XV, ed.
  C.~{Gabriel}, C.~{Arviset}, D.~{Ponz}, \& S.~{Enrique}, 666

\bibitem[{{van Loon} {et~al.}(2005){van Loon}, {Marshall}, \&
  {Zijlstra}}]{vanLoon_etal05}
{van Loon}, J.~T., {Marshall}, J.~R., \& {Zijlstra}, A.~A. 2005, \aap, 442,
  597, \dodoi{10.1051/0004-6361:20053528}

\bibitem[{Wall {et~al.}(1994)}]{perl}
Wall, L., {et~al.} 1994, The Perl programming language,  Prentice Hall Software
  Series

\bibitem[{{Whitmore} {et~al.}(1999){Whitmore}, {Zhang}, {Leitherer}, {Fall},
  {Schweizer}, \& {Miller}}]{whitmore99}
{Whitmore}, B.~C., {Zhang}, Q., {Leitherer}, C., {et~al.} 1999, \aj, 118, 1551,
  \dodoi{10.1086/301041}

\bibitem[{{Williams} \& {Hodge}(2001)}]{williams01}
{Williams}, B.~F., \& {Hodge}, P.~W. 2001, \apj, 548, 190,
  \dodoi{10.1086/318679}

\bibitem[{{Williams} {et~al.}(2014){Williams}, {Lang}, {Dalcanton}, {Dolphin},
  {Weisz}, {Bell}, {Bianchi}, {Byler}, {Gilbert}, {Girardi}, {Gordon},
  {Gregersen}, {Johnson}, {Kalirai}, {Lauer}, {Monachesi}, {Rosenfield},
  {Seth}, \& {Skillman}}]{williams14}
{Williams}, B.~F., {Lang}, D., {Dalcanton}, J.~J., {et~al.} 2014, \apjs, 215,
  9, \dodoi{10.1088/0067-0049/215/1/9}

\bibitem[{{Williams} {et~al.}(2017){Williams}, {Dolphin}, {Dalcanton}, {Weisz},
  {Bell}, {Lewis}, {Rosenfield}, {Choi}, {Skillman}, \&
  {Monachesi}}]{williams17}
{Williams}, B.~F., {Dolphin}, A.~E., {Dalcanton}, J.~J., {et~al.} 2017, \apj,
  846, 145, \dodoi{10.3847/1538-4357/aa862a}

\bibitem[{{Willmer}(2018)}]{willmer18}
{Willmer}, C. N.~A. 2018, \apjs, 236, 47, \dodoi{10.3847/1538-4365/aabfdf}

\end{thebibliography}
\bibliographystyle{aasjournal}



\end{document}